\RequirePackage{fix-cm}
\documentclass[smallextended]{svjour3}       

\include{packages}

\definecolor{v3_1}{RGB}{68, 1, 84}
\definecolor{v3_2}{RGB}{26, 136, 115}
\definecolor{v3_3}{RGB}{177, 162, 25}

\definecolor{v4_1}{RGB}{68, 1, 84}
\definecolor{v4_2}{RGB}{40, 101, 128}
\definecolor{v4_3}{RGB}{75, 161, 78}
\definecolor{v4_4}{RGB}{177, 162, 25}

\definecolor{v5_1}{RGB}{68, 1, 84}
\definecolor{v5_2}{RGB}{52, 81, 130}
\definecolor{v5_3}{RGB}{26, 136, 115}
\definecolor{v5_4}{RGB}{112, 167, 52}
\definecolor{v5_5}{RGB}{177, 162, 25}

\definecolor{v6_1}{RGB}{68, 1, 84}
\definecolor{v6_2}{RGB}{59, 67, 128}
\definecolor{v6_3}{RGB}{32, 116, 125}
\definecolor{v6_4}{RGB}{48, 153, 96}
\definecolor{v6_5}{RGB}{132, 168, 36}
\definecolor{v6_6}{RGB}{177, 162, 25}

\definecolor{v7_1}{RGB}{68, 1, 84}
\definecolor{v7_2}{RGB}{64, 56, 125}
\definecolor{v7_3}{RGB}{40, 101, 128}
\definecolor{v7_4}{RGB}{26, 136, 115}
\definecolor{v7_5}{RGB}{75, 161, 78}
\definecolor{v7_6}{RGB}{144, 167, 27}
\definecolor{v7_7}{RGB}{177, 162, 25}

\definecolor{stdgray}{RGB}{205, 205, 205}
\definecolor{stdyellow}{RGB}{203, 185, 29}
\definecolor{outlinecol}{RGB}{44, 125, 73}

\pgfplotscreateplotcyclelist{viridis3}{
  v3_1,thick,every mark/.append style={thick,fill=v3_1},mark=o\\
  v3_2,thick,every mark/.append style={thick,fill=v3_2},mark=square*\\
  v3_3,thick,every mark/.append style={thick,fill=v3_3},mark=diamond\\
}

\pgfplotscreateplotcyclelist{viridis4}{
  v4_1,thick,every mark/.append style={thick,fill=v4_1},mark=*\\
  v4_2,thick,every mark/.append style={thick,fill=v4_2},mark=square*\\
  v4_3,thick,every mark/.append style={thick,fill=v4_3},mark=diamond*\\
  v4_4,thick,every mark/.append style={thick,fill=v4_4},mark=triangle*\\
}

\pgfplotscreateplotcyclelist{viridis5}{
  v6_1,thick,every mark/.append style={thick,fill=v6_1},mark=o\\
  v6_2,thick,every mark/.append style={thick,fill=v6_2},mark=square*\\
  v6_3,thick,every mark/.append style={thick,fill=v6_3},mark=diamond\\
  v6_4,thick,every mark/.append style={thick,fill=v6_4},mark=triangle*\\
  v6_5,thick,every mark/.append style={thick,fill=v6_5},mark=star\\
}

\pgfplotscreateplotcyclelist{correl}{
  color=black,mark=none\\
  v6_1,thick,every mark/.append style={thick,fill=v6_1},mark=triangle*\\
  v6_2,thick,every mark/.append style={thick,fill=v6_2},mark=*\\
  v6_3,thick,every mark/.append style={thick,fill=v6_3},mark=star\\
  v6_4,thick,every mark/.append style={thick,fill=v6_4},mark=triangle\\
  v6_5,thick,every mark/.append style={thick,fill=v6_5},mark=o\\
}

\pgfplotscreateplotcyclelist{viridis6}{
  v6_1,thick,every mark/.append style={thick,fill=v6_1},mark=o\\
  v6_2,thick,every mark/.append style={thick,fill=v6_2},mark=square*\\
  v6_3,thick,every mark/.append style={thick,fill=v6_3},mark=diamond\\
  v6_4,thick,every mark/.append style={thick,fill=v6_4},mark=triangle*\\
  v6_5,thick,every mark/.append style={thick,fill=v6_5},mark=star\\
  v6_6,thick,every mark/.append style={thick,fill=v6_6},mark=*\\
}

\pgfplotscreateplotcyclelist{viridis62}{
  v6_2,thick,every mark/.append style={thick,fill=v6_2},mark=o\\
  v6_3,thick,every mark/.append style={thick,fill=v6_3},mark=square*\\
  v6_4,thick,every mark/.append style={thick,fill=v6_4},mark=diamond\\
  v6_5,thick,every mark/.append style={thick,fill=v6_5},mark=triangle*\\
  v6_6,thick,every mark/.append style={thick,fill=v6_6},mark=star\\
}

\smartqed  

\usepackage{tikz}
\usepackage{xifthen}
\usetikzlibrary{spy,arrows}
\usetikzlibrary{arrows.meta}
\usetikzlibrary{shapes.geometric}
\usetikzlibrary{intersections}
\usetikzlibrary{positioning}
\usetikzlibrary{backgrounds}
\usepackage{pgfplots}
\pgfplotsset{compat=1.13}
\usetikzlibrary{calc}
\tikzstyle{rounded}  = [rounded corners]     
\tikzstyle{function} = [rectangle, text centered, draw=black,minimum width=6.0cm,minimum height=1.0cm]
\tikzstyle{mpi}      = [anchor=south west, xshift=0.2cm] 

\usetikzlibrary{3d}

\usetikzlibrary{fillbetween}
\usetikzlibrary{decorations,decorations.text,backgrounds}
\usetikzlibrary{fadings}
\usetikzlibrary{shapes.arrows}

\newcounter{tikzsubfigcounter}[figure]
\renewcommand{\thetikzsubfigcounter}{\the\numexpr\value{figure}+1\relax\alph{tikzsubfigcounter}}

\newcommand{\tikztitle}[1]{ %
  \refstepcounter{tikzsubfigcounter}
  \textbf{(\alph{tikzsubfigcounter})}\space\space #1 
}

\newcounter{tikzsubfigcounterinvisible}[figure]
\renewcommand{\thetikzsubfigcounterinvisible}{\the\numexpr\value{chapter}.\the\numexpr\value{figure}+1\relax(\alph{tikzsubfigcounterinvisible})}

\usepgfplotslibrary{groupplots}

\newlength{\figureheight}
\newlength{\figurewidth}
\newcommand{\pounce}{\textsc{POUNCE}\xspace}

\DeclareMathAlphabet{\mathesstixfrak}{U}{esstixfrak}{m}{n}

\newcommand{\R}{\mathbb{R}}
\newcommand{\N}{\mathbb{N}}

\renewcommand{\vec}[1]{{\bm{#1}}} 
\newcommand{\uSym}{u}
\newcommand{\U}{\vec{\uSym}}

\newcommand{\xSym}{x}
\newcommand{\x}{\vec{\xSym}}
\newcommand{\tsym}{{t}}
\DeclareMathOperator*{\eop}{end}
\newcommand{\tEnd}{\tsym_{\eop}}
\newcommand{\ddt}{\partial_{\tsym}}

\newcommand{\gradx}{{\vec{{\nabla}}_{\x}}}
\newcommand{\divx}{\gradx\cdot}
\newcommand{\xref}{\vec{\eta}}
\newcommand{\gradxref}{{\vec{{\nabla}}_{\xref}}}
\newcommand{\divxref}{\gradxref\cdot}
\newcommand{\fluxconv}{\vec{G}}
\newcommand{\fluxvisc}{\vec{H}}
\newcommand{\fluxboth}{\vec{F}}
\newcommand{\CVfluxboth}{\pmb{\mathcal{F}}}

\newcommand{\physdom}{D}
\DeclareMathOperator*{\varop}{var}
\newcommand{\nvar}{{n_{\varop}}}
\newcommand{\ndim}{3}

\newcommand{\jacd}{\mathesstixfrak{J}}
\newcommand{\refelem}{E}
\newcommand{\testfunc}{\phi}
\newcommand{\ndg}{N}
\newcommand{\normal}{\vec{\mathfrak{n}}}

\newcommand{\dens}{{\rho}}
\newcommand{\vel}{{v}}
\newcommand{\mom}{\mathesstixfrak{m}}
\newcommand{\velv}{{\vec{\vel}}}
\newcommand{\ener}{{e}}
\newcommand{\pres}{{p}}
\newcommand{\stress}{{\tau}}
\newcommand{\heatflux}{\mathfrak{Q}}
\newcommand{\visc}{{\mu}}
\newcommand{\heatcond}{{\lambda}}
\newcommand{\isencoef}{{\gamma}}

\newcommand{\tempr}{{\mathcal{T}}}
\newcommand{\gk}{{R}}

\DeclareMathOperator{\E}{\mathbb{E}}
\DeclareMathOperator{\Var}{\mathbb{V}ar}
\DeclareMathOperator{\dE}{E}
\DeclareMathOperator{\dVar}{Var}

\DeclareMathOperator{\dSD}{SD}
\newcommand{\expec}{\mu}
\newcommand{\estim}[1]{\widetilde{#1}}

\newcommand{\stddev}{\sigma}

\newcommand{\samplespace}{{\Omega}}
\newcommand{\stochdom}{{\Gamma}}

\newcommand{\xicmp}{{\xi}}
\newcommand{\xiv}{{\vec{\xicmp}}}
\newcommand{\ndimstoch}{{J}}
\newcommand{\stochdimind}{{j}}
\newcommand{\pdfsym}{{\varrho}}
\newcommand{\pdf}{{\pdfsym_{\stochdom}(\xiv)}}

\newcommand{\sAlgebra}{\mathcal{F}}
\newcommand{\pMeasure}{\mathbb{P}}
\newcommand{\iSample}{m}
\newcommand{\nSamples}{M}
\newcommand{\scqoi}{q}
\newcommand{\sdqoi}{\scqoi}

\newcommand{\xiunif}{\xi} 
\newcommand{\unifLB}{a}
\newcommand{\unifUB}{b}

\newcommand{\basis}{{\Psi}}

\newcommand{\coef}[1]{{\widehat{#1}}}

\newcommand{\varg}{{g}}
\newcommand{\varh}{{h}}

\newcommand{\stochind}{{\kappa}}
\DeclareMathOperator*{\stoch}{stoch}

\newcommand{\maxstochpolydeg}{N^{\stoch}}
\newcommand{\indspace}{{\mathcal{K}}}

\newcommand{\xiQuadIndex}{\ensuremath{\beta}}
\newcommand{\Pidxset}{B}
\newcommand{\nQP}{{\ensuremath{n}_{\xiQuadIndex}}}

\newcommand{\quadRweight}[1]{{\omega}_{#1}}

\newcommand{\secref}[1]{Section~\ref{#1}}
\newcommand{\figref}[1]{Figure~\ref{#1}}
\newcommand{\tabref}[1]{Table~\ref{#1}}

\newcommand{\equref}[1]{\eqref{#1}}

\newcommand{\uniformdist}{\mathcal{U}}

\newcommand{\err}{\mathcal{E}}

\DeclareSymbolFont{mathroman}{OT1}{stix}{m}{it}
\DeclareSymbolFontAlphabet{\mathroman}{mathroman}

\newcommand{\levelind}{\:\!\mathroman{l}}

\newcommand{\nlevels}{L}
\DeclareMathOperator*{\MLMCop}{MLMC}
\newcommand{\dEMLMC}{\dE_{\MLMCop}}
\newcommand{\dVarMLMC}{\dVar_{\MLMCop}}
\newcommand{\levelsum}{\sum_{\levelind=1}^{\nlevels}}

\newcommand{\work}{\mathcal{W}}

\newcommand{\cBudget}{\mathcal{B}}

\DeclareMathOperator*{\opt}{opt}

\newcommand{\mlopt}{\nSamples^{\levelind}_{\opt}}

\DeclareMathOperator*{\MFMCop}{MFMC}

\newcommand{\iModel}{\levelind}
\newcommand{\nModels}{\nlevels}
\newcommand{\dEMFMC}{\dE_{\MFMCop}}

\DeclareMathOperator*{\HF}{HF}

\newcommand{\modelsum}[1]{\sum_{\iModel=#1}^{\nModels}}
\newcommand{\mfmcAl}{\alpha}
\newcommand{\ds}{\displaystyle}
\newcommand{\pears}{\mathcal{R}}

\DeclareMathOperator*{\elemOp}{el}
\newcommand{\nElem}{n_{\elemOp}}

\newcommand{\cp}{C_{p}}
\newcommand{\cl}{c_{l}}
\newcommand{\cd}{c_{d}}
\newcommand{\chord}{c}
\DeclareMathOperator*{\aoa}{AoA}
\DeclareMathOperator*{\mach}{Ma}
\DeclareMathOperator*{\reynolds}{Re}

\newcommand{\velInf}{\vel_{\infty}}

\newcommand{\dzCav}{d_{z}}

\newcommand{\dxp}{\Delta x^{+}}
\newcommand{\dyp}{\Delta y^{+}}
\newcommand{\dzp}{\Delta z^{+}}

\newcommand{\degC}{{}^{\circ}\text{C}}

\newcommand{\pcaiIn}{s}
\newcommand{\pcaVecIn}{\vec{a}}
\DeclareMathOperator*{\Fop}{F}
\newcommand{\pcaNFeat}{n_{\Fop}}
\newcommand{\pcaNSamples}{n_{\pcaiIn}}
\newcommand{\pcaMatIn}{\vec{\mathcal{A}}}
\newcommand{\pcaPC}{\vec{U}}
\newcommand{\pcaPCcol}{\mathfrak{u}}
\newcommand{\pcaSVMat}{\vec{{\Sigma}}}
\newcommand{\pcaSV}{\mathesstixfrak{s}}
\newcommand{\pcaLS}{\vec{V}}

\newcommand{\signedD}{d_{+}}
\newcommand{\dmax}{d_{\max}}
\newcommand{\signSym}{\mathcal{S}}
\DeclareMathOperator*{\inop}{in}
\newcommand{\modeFunc}{d_{\inop}}
\DeclareMathOperator*{\outop}{out}
\newcommand{\pcaOut}{d_{\outop}}

\newcommand{\lift}{F_{l}}
\DeclareMathOperator*{\dynop}{dyn}
\newcommand{\dynpres}{\pres_{\dynop}}
\newcommand{\drag}{F_{d}}
\newcommand{\dzAF}{\dzCav}

\journalname{International Journal for Numerical Methods in Engineering}
\begin{document}

\title{Data-integrated uncertainty quantification for the performance prediction of iced airfoils}

\titlerunning{UQ for iced airfoil performance}        

\author{Jakob Dürrwächter         \and
        Andrea Beck               \and 
        Claus-Dieter Munz
}


\institute{J. Duerrwaechter \at
              Institute for Aerodynamics and Gas Dynamics, University of Stuttgart \\
              \email{jd@iag.uni-stuttgart.de}           
           \and
           A. Beck \at
              Institute for Aerodynamics and Gas Dynamics, University of Stuttgart \\
              \email{beck@iag.uni-stuttgart.de}           
           \and
           C.-D. Munz \at
              Institute for Aerodynamics and Gas Dynamics, University of Stuttgart
}

\date{Received: date / Accepted: date}

\maketitle

\begin{abstract}
  Airfoil icing is a severe safety hazard in aviation and causes power losses on wind turbines. The precise shape of the ice formation is subject to large uncertainties, so uncertainty quantification (UQ) is needed for a reliable prediction of its effects. In this study, we aim to establish a reliable estimate of the effect of icing on airfoil performance through UQ. We use a series of experimentally measured wind tunnel ice shapes as input data. Principal component analysis is employed to construct a set of linearly uncorrelated geometric modes from the data, which serves as random input to the UQ simulation. For uncertainty propagation, non-intrusive polynomial chaos expansion (NIPC), multi-level Monte Carlo (MLMC) and multi-fidelity Monte Carlo control variate (MFMC) methods are employed and compared. As a baseline model, large eddy simulations (LES) are carried out using the discontinuous Galerkin flow solver FLEXI. UQ simulations are carried out with the in-house framework \pounce (Propagation of Uncertainties). Its focus is on a high level of automation and efficiency considerations in a high performance computing environment. Due to the high number of samples, the simulation tool chain of the baseline model is completely automatized, including a new structured boundary layer grid generator for highly irregular domain shapes. Results show that forces on the airfoil vary considerably due to the uncertain ice shape. All three methods prove to be suited to predict mean and standard deviation. In the Monte Carlo techniques, the choice and performance of low-fidelity models is shown to be decisive for estimator variance reduction. The MFMC method performs best in this study. To our knowledge, there are no UQ-studies of iced airfoils based on LES, let alone with advanced UQ methods such as MLMC or MFMC. The present study thus represents a leap in accuracy and level of detail for this application. 

\keywords{Computational fluid dynamics \and Uncertainty quantification \and Airfoil icing \and Monte Carlo \and discontinuous Galerkin \and High performance computing \and Automated grid generation}
\subclass{MSC 76F65 \and MSC 65C05} 

\end{abstract}

\section{Introduction}
\label{sec:intro}

Airfoil icing is a serious problem in aviation and wind energy. In aviation, the consequences can be fatal. Recent examples are the crashes of Aero Caribbean Flight 883 in 2010 and Sol Líneas Aéreas Flight 5428 in 2011. 
In wind energy, blade icing reduces the yearly energy production up to at least 17 \% \cite{Barber2011}. In Europe, 20 \% of wind turbines are built in areas where icing has to be taken into account \cite{Tammelin1998,Tammelin2000}.
The prediction of iced airfoil performance helps to avoid icing conditions by planning different flight envelopes and wind turbine sites, and it helps to design anti-icing and deicing mechanisms.

Airfoil icing is typically caused by super-cooled liquid droplets which impinge on the airfoil leading edge and freeze. Two basic types can be distinguished: 
\textit{Rime ice} is white and opaque, with a rough surface. It is formed in temperatures significantly below the freezing point ($<-15\degC$ according to \cite{Janjua2018}) by small supercooled droplets, which freeze instantly on impact on the airfoil surface. Rime ice accretes in rather regular shapes as a layer around and downstream of the leading edge. Its aerodynamic effects are mostly moderate, and can include laminar-to-turbulent transition triggered by the roughness or small separation bubbles at the end of the ice layer.  
\textit{Glaze ice} is transparent, with a relatively smooth surface. It is formed in temperatures close to the freezing point. Here, the ice forms a liquid film on the airfoil surface, which may run downstream on the surface driven by the surrounding flow before freezing. The macroscopic shapes tend to be more irregular: Large ice horns typically protrude from the leading edge upstream into the flow, on the pressure side, suction side, or both. The aerodynamic effect is major, with large separation bubbles behind the horns and a heavily impaired flow field. Other types of icing include a mixed form of rime and glaze, or supercooled-large droplet ice, which is a rare, but very impactful form of icing. Further details on ice shapes and their effect on airfoil flow fields and airfoil performance are given in the review by Bragg \cite{Bragg2005}. 

The effects of icing can be predicted numerically. The flow fields are highly disturbed, depend strongly on the exact location of laminar-to-turbulent transition as well as flow separation behind the iced region. Reynolds-averaged Navier-Stokes (RANS) simulations, which are used for most airfoil flow simulations, perform poorly in predicting these phenomena. Instead, scale-resolving simulation such as large eddy simulations (LES) are needed to accurately capture the formation and effect of turbulent eddies under these special circumstances. A review of previous numerical studies of iced airfoils is given in \cite{Stebbins2019}, which includes some scale-resolving simulations, mostly detached eddy simulations. The work \cite{Brown2013} is an early example of an iced airfoil LES.
These studies, however, are deterministic and do not consider uncertainties due to varying ice shapes.

Turbulence is a multi-scale phenomenon. Numerical methods with a high order of accuracy are especially efficient for these types of problems. In this study, the discontinuous Galerkin method is used, which is implemented in our in-house open-source solver FLEXI  \cite{Krais2019}. The resolution requirements of LES entail the need for high performance computing resources. FLEXI achieves excellent scaling on such massively parallel clusters.

Ice on airfoils appears in various shapes. On the one hand, this is due to different icing conditions, such as differences in temperature, mean droplet diameter or liquid water content in the air. Random fluctuations in droplet density or turbulence in the flow also lead to random variations, as can be seen, for example, in spanwise variations of the ice shape along a wing. Which ice shape occurs on a wing is thus uncertain. In a numerical simulation, this translates to an uncertain simulation setup. Neglecting these uncertainties and considering one ice shape entails limited predictive capabilities. However, uncertain parameters can be taken into account in numerical simulations and their effect can be quantified. This is called uncertainty quantification (UQ).

In UQ, some of the input parameters are assumed to be uncertain with a given random distribution. Since the output quantity of interest is a function of this input, it is also random. The objective of forward UQ is to estimate properties of the random distribution of the output, such as its mean and variance. Most methods rely on several evaluations of a standard model with different realizations of the random input parameters. This is called non-intrusive uncertainty quantification. A large number of non-intrusive UQ methods have been developed, especially in recent years, but their application to complex engineering problems is still rare, and research is still needed about their performance in such practical problems.

In this paper, the effect of airfoil icing with an uncertain ice geometry is investigated using different UQ methods. The ice accretion process itself is not modeled. Instead, a data-driven approach is taken: A set of experimentally measured ice shapes is used as simulation input. A continuous random vector is generated from this data set with a principal component analysis. Three non-intrusive UQ methods are employed for uncertainty propagation, to combine the advantages of each and to allow for a comparison of the methods: The non-intrusive polynomial chaos (NIPC) method, the multilevel Monte Carlo (MLMC) method and the multifidelity Monte Carlo (MFMC) method. Due to the high number of samples in Monte Carlo methods, all pre-and post-processing steps of the baseline computational setup are completely automatized. This includes structured boundary layer grid generation with random ice shapes, for which a new algorithm was developed.

There are a few previous UQ studies on iced airfoil flow, presented in \cite{DeGennaroHeuristic2015,DeGennaroData2015} and \cite{Tabatabaei2019}, as well as one simulating the ice accretion process under uncertain conditions \cite{Gori2022}. The idea to use principal component analysis to create a low-dimensional parametrization of ice shapes was first presented in \cite{DeGennaroData2015} and the present paper builds on this study regarding this methodology. However, there are several notable differences and novelties in the present paper compared to this and other previous studies: In all cited studies, a two-dimensional Reynolds-averaged Navier-Stokes (RANS) solver serves as a baseline model, where much more accurate large-eddy simulations are employed in the present work. Furthermore, in all previous studies, only the NIPC method is used, while several UQ methods including MLMC and MFMC are compared here. Moreover, compared with \cite{DeGennaroData2015}, a different input data set is used and the input data parametrization method prior to the principal component analysis is different. 

The main novelty of the present work is thus the combination of various advanced methods in different the areas of numerics and UQ, their efficient implementation in the environment of high performance computing, and the application to the complex problem of airfoil icing. This new combination allows insights in different areas of research: On the physical side, the effects of airfoil icing are predicted in a holistic view. Additionally, the study includes development and comparative investigation of the UQ methods, with respect to their implementation in high performance systems (which is realized with a dedicated in-house software package \pounce \cite{Duerrwaechter2023}) and their performance under real-world circumstances. This makes the findings of the paper not only relevant to airfoil icing, but to prospective uses of UQ in other fluid dynamics problems. 

The paper is structured as follows. \secref{sec:cfd} introduces the baseline model, including the governing equations and their discretization with the discontinuous Galerkin method, a reference to the in-house software FLEXI, an automated grid generation algorithm developed for this study, and parameter setup of the numerical model.
\secref{sec:uq} covers the UQ aspect, including a description of the input ice shape parametrization using principal component analysis, the employed UQ methods, the chosen method parameters, and implementation of the methods on an HPC system using the in-house framework \pounce. Results are presented and discussed in \secref{sec:results}. The paper is summarized and an outlook is provided in \secref{sec:conclusion}.

\section{Baseline computational fluid dynamics model}
\label{sec:cfd}

In this chapter, the baseline deterministic model is discussed. \secref{sec:nse} introduces the governing equations. \secref{sec:dgsem} introduces the numerical methods which are employed in the flow solver FLEXI. FLEXI serves as a deterministic baseline code for all simulations presented in the following and is described in \secref{sec:flexi}. \secref{sec:grid_generation} describes the automated grid generation algorithm, \secref{sec:cfd_prms} considers model setup and parameters.

\subsection{The compressible Navier-Stokes equations}\label{sec:nse}

The compressible Navier-Stokes equations are the governing equations for the description of the flow of a viscous, compressible, Newtonian fluid. They are conservation laws of mass, momentum and energy. In conservation form, they read
\begin{equation} \label{eq:ConservationLaw}
  \ddt \U + \divx (\fluxconv(\U)-\fluxvisc(\U,\gradx\U))=0 \quad \text{in } \physdom\times\R_{+},
\end{equation}
where $\U = \U(\vec{x},\tsym) \,:\,\physdom\times\R_{+} \rightarrow \mathcal{U}\subset \R^\nvar $ denotes the vector of conserved quantities and $\ddt$ represents the partial derivative with respect to time, $\gradx$ is the gradient operator in physical space $\physdom\subset\mathbb{R}^\ndim$ assuming a three-dimensional problem. $\fluxconv$ and $\fluxvisc$ describe the convective  and viscous flux tensors, respectively, which are summarized for brevity as $\fluxboth(\U,\gradx\U)\coloneqq \fluxconv(\U) + \fluxvisc(\U,\gradx\U)$.
The deterministic conserved quantities are 
\begin{equation} \label{eq:state_vector}
  \U = (\dens, \mom_1, \mom_2, \mom_3, \dens\ener)^T = (\dens, \dens\vel_1, \dens\vel_2, \dens\vel_3, \dens\ener)^T, 
\end{equation}
where $\dens$ is the mass density, $\mom_i$ and $\vel_i$ are the momentum and velocity in $i$-direction, respectively, and $\ener$ is the stagnation energy. The $i$th column of the convective and viscous fluxes is given by 
\begin{equation} \label{eq:fluxes}
  \fluxconv^i = 
  \begin{pmatrix} 
    \dens\vel_i \\ 
    \dens\vel_1\vel_i + \delta_{1i}\pres \\ 
    \dens\vel_2\vel_i + \delta_{2i}\pres \\ 
    \dens\vel_3\vel_i + \delta_{3i}\pres \\ 
    \dens\ener\vel_i + \pres\vel_i \\ 
  \end{pmatrix} \quad\text{and}\quad
  \fluxvisc^i = 
  \begin{pmatrix} 
    0 \\ 
    \stress_{1i} \\ 
    \stress_{2i} \\ 
    \stress_{3i} \\ 
    \stress_{ij}\vel_j - \heatflux_i \\ 
  \end{pmatrix}, \qquad i=1,2,3.
\end{equation}
Here, $\delta_{ij}$ is the Kronecker delta and $\pres=\pres(\U)$ is the pressure given by the ideal gas equation
\begin{equation} \label{eq:pressure}
  \pres = \dens\gk\tempr = (\isencoef-1)\dens\left(\ener - \frac{1}{2} \velv\cdot\velv\right)
\end{equation}
with temperature $\tempr$. Furthermore, $\stress_{ij}$ denote the entries of the viscous stress tensor
\begin{equation}
  \vec{\stress} = \visc\left(\gradx\velv + (\gradx\velv)^T - \frac{2}{3}(\divx\velv)\vec{I}\right), 
\end{equation}
and $\heatflux_i$ are the components of the heat flux vector $\vec{\heatflux}=-\heatcond\gradx\tempr$. The fluid-dependent variables used above are the specific gas constant $\gk$, the adiabatic coefficient $\isencoef$, the dynamic viscosity $\visc$ and the heat conductivity $\heatcond$. They are all assumed to be constant in space and time. The adiabatic coefficient is chosen as $\isencoef=1.4$ throughout this work. 

\subsection{The discontinuous Galerkin spectral element method}\label{sec:dgsem}

A suitable numerical scheme is to be chosen for discretization of the governing equations in space and time. 
Methods with a high order of accuracy promise crucial efficiency gains for multiscale problems, such as the scale-resolving simulations of turbulent flows carried out in this work. Discontinuous Galerkin (DG) methods are a prominent class of high-order methods.
In this work, the discontinuous Galerkin spectral element method (DGSEM) in combination with explicit Runge-Kutta time-stepping is used for the space-time discretization of the Navier-Stokes equations and other conservation laws. 
This section only provides a short summary of the method, further details can be found for example in \cite{Krais2019,Kopriva2009,Hesthaven2008,Bassi1997}.

The physical domain $\physdom$ is partitioned into $\nElem \in \N$ disjoint curved hexahedral elements. The basic idea of DG is to discretize each of these elements with the Galerkin method and to allow for discontinuities between the elements. The numerical fluxes between the grid elements are determined by use of Riemann solvers. In the DGSEM, the solution on each element is approximated by a tensor product of polynomials, which are expressed in a Lagrange basis.

For reducing computational complexity, all elements are transformed from physical space onto the reference element $\refelem(\vec{\xref})=[-1,1]^\ndim$, such that for each element the governing equations become
\begin{equation} 
  \ddt \jacd \U + \divxref \CVfluxboth(\U,\gradx\U)=0,
\end{equation}
where $\jacd$ is the Jacobian determinant of the mapping $\x(\xref)$ and $\CVfluxboth$ is the so-called contravariant flux \cite{Kopriva2009}. Multiplication with a test function $\testfunc$ and integration over the reference element $\refelem$ yields
\begin{equation}
   \ddt \int_{\refelem} \jacd  \U   \testfunc\, d\xref  + \int_{\refelem} \divxref\CVfluxboth(\U,\gradx\U) \testfunc\, d\xref =0. 
\end{equation}
This corresponds to a projection onto the space spanned by the test functions. 
Integration by parts leads to the weak form of \equref{eq:ConservationLaw}:
\begin{equation}\label{eq:dg}
  \ddt \int_{\refelem} \jacd  \U  \testfunc\, d\xref  + \oint_{\partial\refelem} (\CVfluxboth \cdot \normal)^* \testfunc\, d\xref  
     - \int_{\refelem} \CVfluxboth(\U,\gradx\U) \cdot \gradxref\testfunc\, d\xref =0, 
\end{equation}
where $\normal$ is the outward pointing unit normal vector of the reference element surface. At element interfaces, discontinuities are allowed, such that the flux $\CVfluxboth$ is not uniquely defined. Riemann solvers are used to approximate the convective flux and the arithmetic mean of element traces is employed for the viscous flux. This approximation is indicated by the superscript $^*$. Components of the solution vector are approximated on each three-dimensional element by a tensor product of one-dimensional polynomials of order $\ndg\in \N$. Lagrange polynomials are employed as both basis and test functions. A collocation approach following \cite{Kopriva2009} using the \mbox{$(\ndg+1)^\ndim$} Gauss-Legendre
or Legendre-Gauss-Lobatto quadrature points is used for interpolation and integration.

In \equref{eq:ConservationLaw}, the gradient of the solution $\gradx\U$ is needed as an input to the viscous terms of the flux function. In the context of a DG discretization, this gradient is obtained with a procedure called lifting as described in \cite{Bassi1997}. To this end, \equref{eq:ConservationLaw} is re-written into an extended system of first-order equations, which are solved sequentially in each time step with the DGSEM method.

To advance the solution in time, a high-order explicit Runge-Kutta time integration scheme of Carpenter\&Kennedy \cite{carpenter1994fourth} is used, which is implemented in a low storage version.
The time step size is constrained by a Courant-Friedrichs-Lewy (CFL) type condition.

In order to improve the robustness of the method, a split form variant of the DGSEM presented in \cite{Gassner2016} is used. It exploits the summation-by-parts property of the DGSEM volume integral operator on Legendre-Gauss-Lobatto nodes and introduces a two-point flux function in the volume integral to achieve properties like the preservation of entropy or kinetic energy. In this work, the kinetic-energy-preserving two-point flux from Pirozzoli \cite{Pirozzoli2011} is used.

\subsection{Baseline flow solver: FLEXI}
\label{sec:flexi}

The methods described above are implemented in the in-house flow solver FLEXI, which is open source\footnote{\texttt{www.flexi-project.org}} and described in detail in \cite{Krais2019}. It can handle unstructured hexahedral grids and is optimized for efficient massively parallel runs, with perfect scaling up to 500.000 cores. It has been successfully applied to various real-world engineering problems, such as cavity aeroacoustics \cite{Kuhn2019} or turbine flow \cite{Kopper2021}. For further details, refer to \cite{Krais2019}. The adapted version used in this work is also open source\footnote{\texttt{https://github.com/flexi-framework/flexi-extensions/tree/pounce_ice/}}.

\subsection{Automated grid generation}
\label{sec:grid_generation}

Structured grids are preferred for boundary layer treatment in LES and are thus used here as well. The DGSEM requires curved hexahedral elements. Elements must not be bent or distored too strongly, as that leads to large non-linear terms in the polynomial geometry mapping, which entails excessive geometric aliasing errors. In the MFMC simulations in this work, thousands of samples are calculated. A grid has to be generated for each sample. Manual grid generation is thus impractical. The grid generation has to be fully automated and highly robust. 

The algorithm for the structured boundary layer grid generation developed for this study is described briefly in Appendix \ref{sec:grid_appendix} and open source\footnote{\texttt{https://github.com/flexi-framework/flexi-extensions/blob/\allowbreak pounce_ice/\allowbreak tools/\allowbreak generate_ice_shapes_and_mesh/make_mesh.py}}. It is based on Laplace smoothing amended with additional terms to ensure grid quality constraints. A previous approach based on radial basis functions did not yield satisfactory results, as the method does not adapt to locally altered wall normals, which results in invalid grids for the investigated heavily iced geometries.

\begin{figure}
  \centering
  \setlength{\fboxsep}{0pt}%
  \setlength{\fboxrule}{0.5pt}%
  \tikztitle{Leading edge ice shape samples}\\\vspace*{0.1cm}
  \fbox{\includegraphics[width=4.2cm, trim=55mm 65mm 30mm 55mm,clip=true]{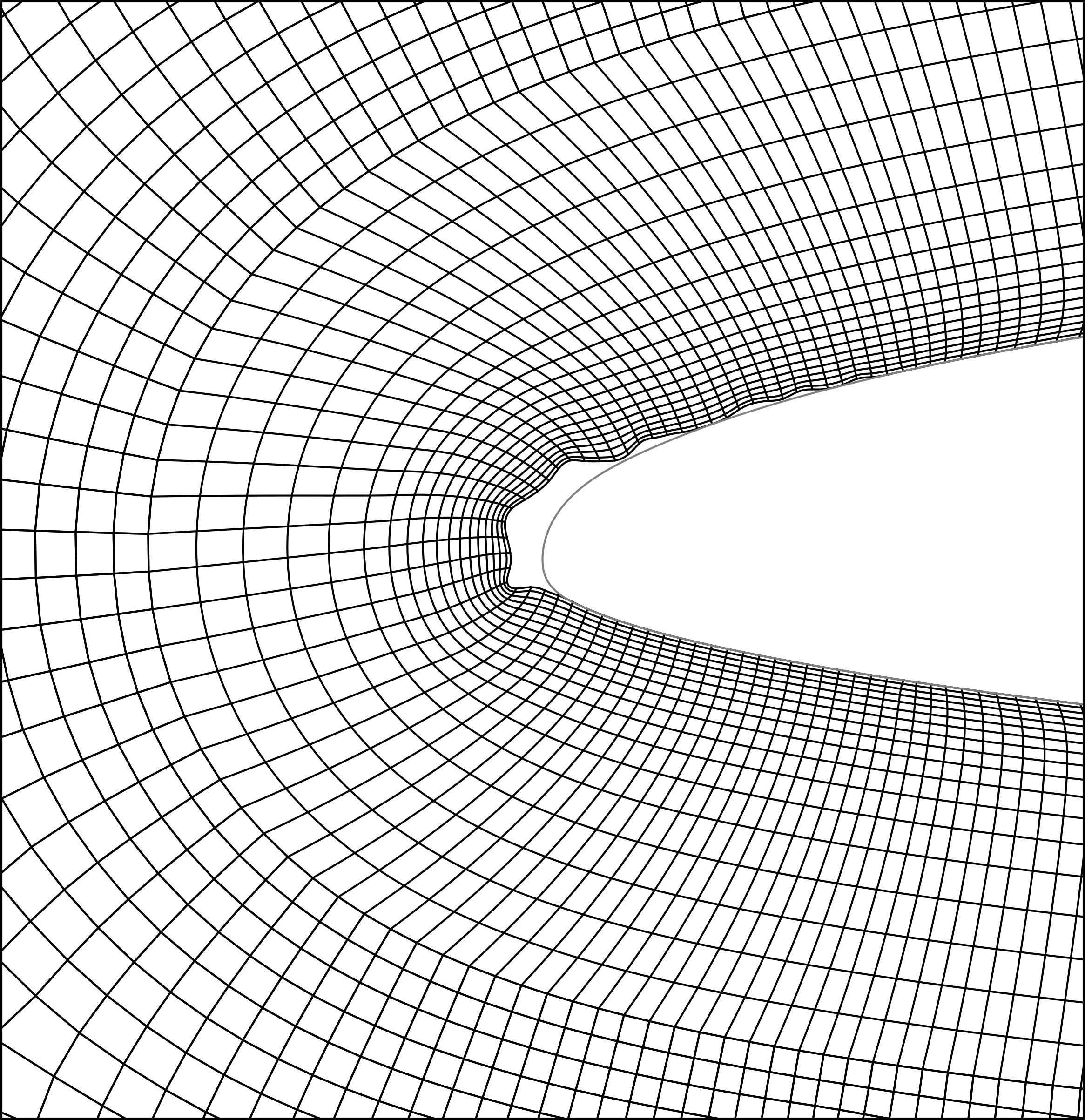}}\hspace{3mm}
  \fbox{\includegraphics[width=4.2cm, trim=55mm 65mm 30mm 55mm,clip=true]{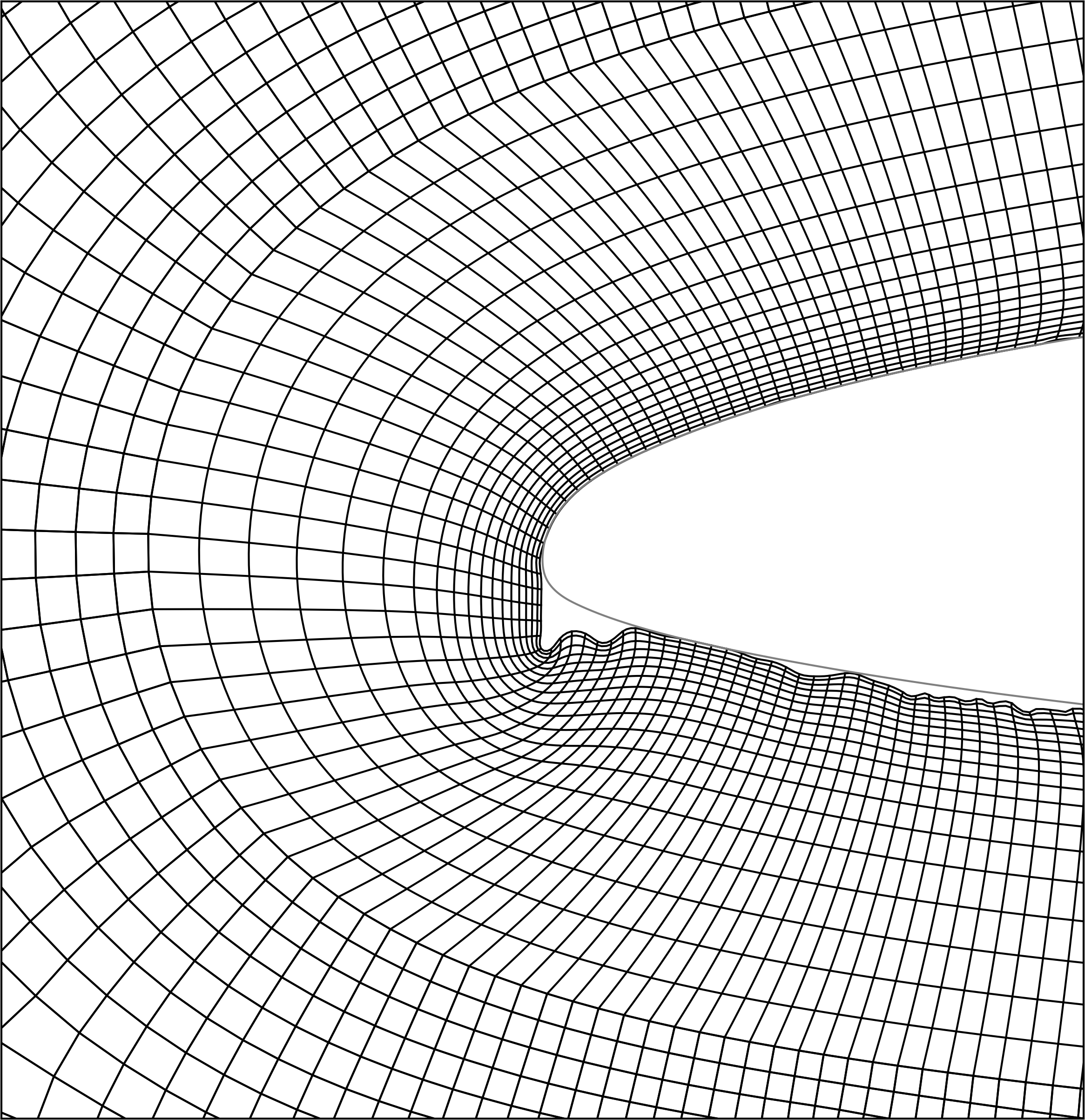}}\\\vspace*{4mm}
  \fbox{\includegraphics[width=4.2cm, trim=55mm 65mm 30mm 55mm,clip=true]{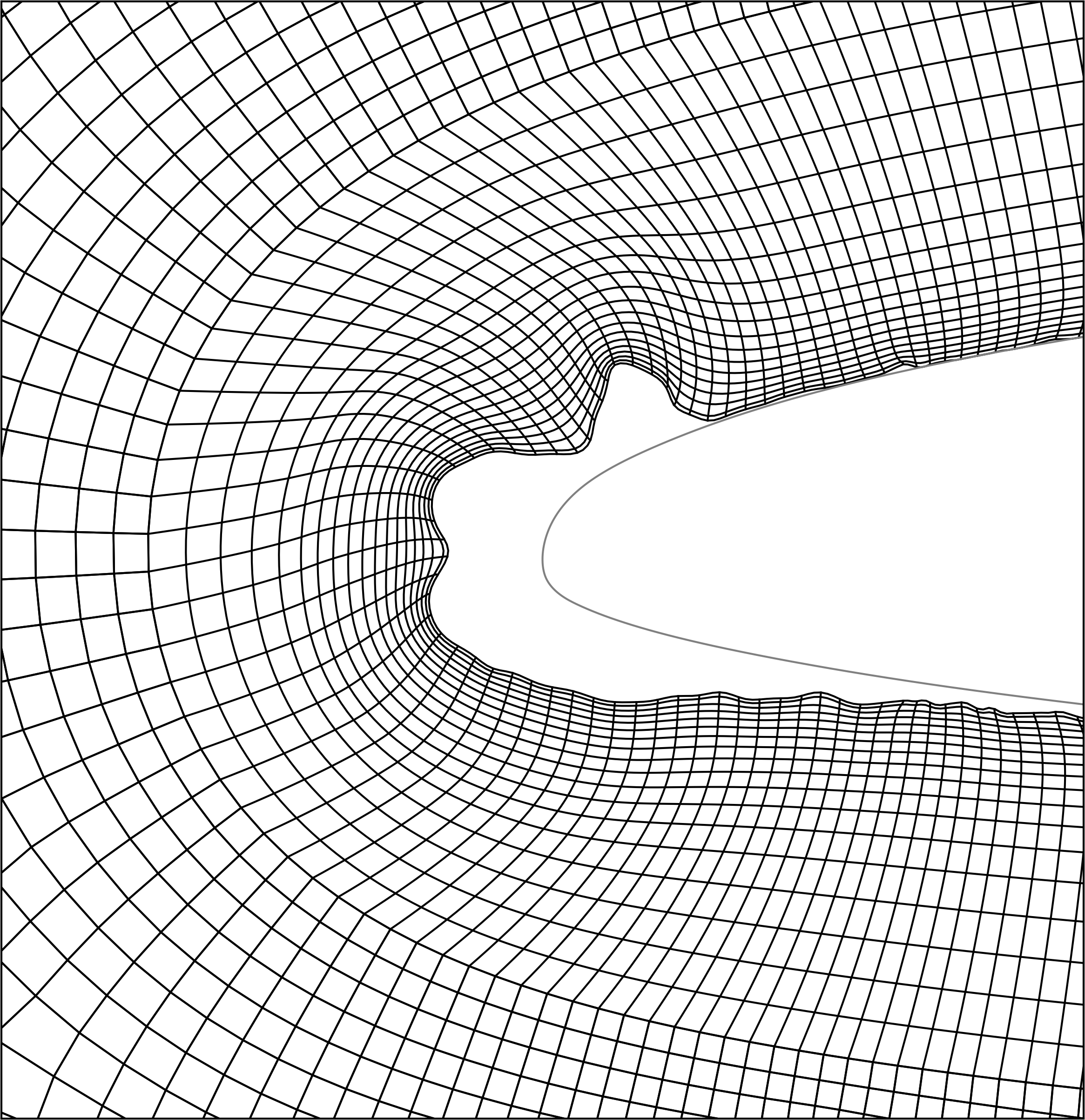}}\hspace{3mm}
  \fbox{\includegraphics[width=4.2cm, trim=55mm 65mm 30mm 55mm,clip=true]{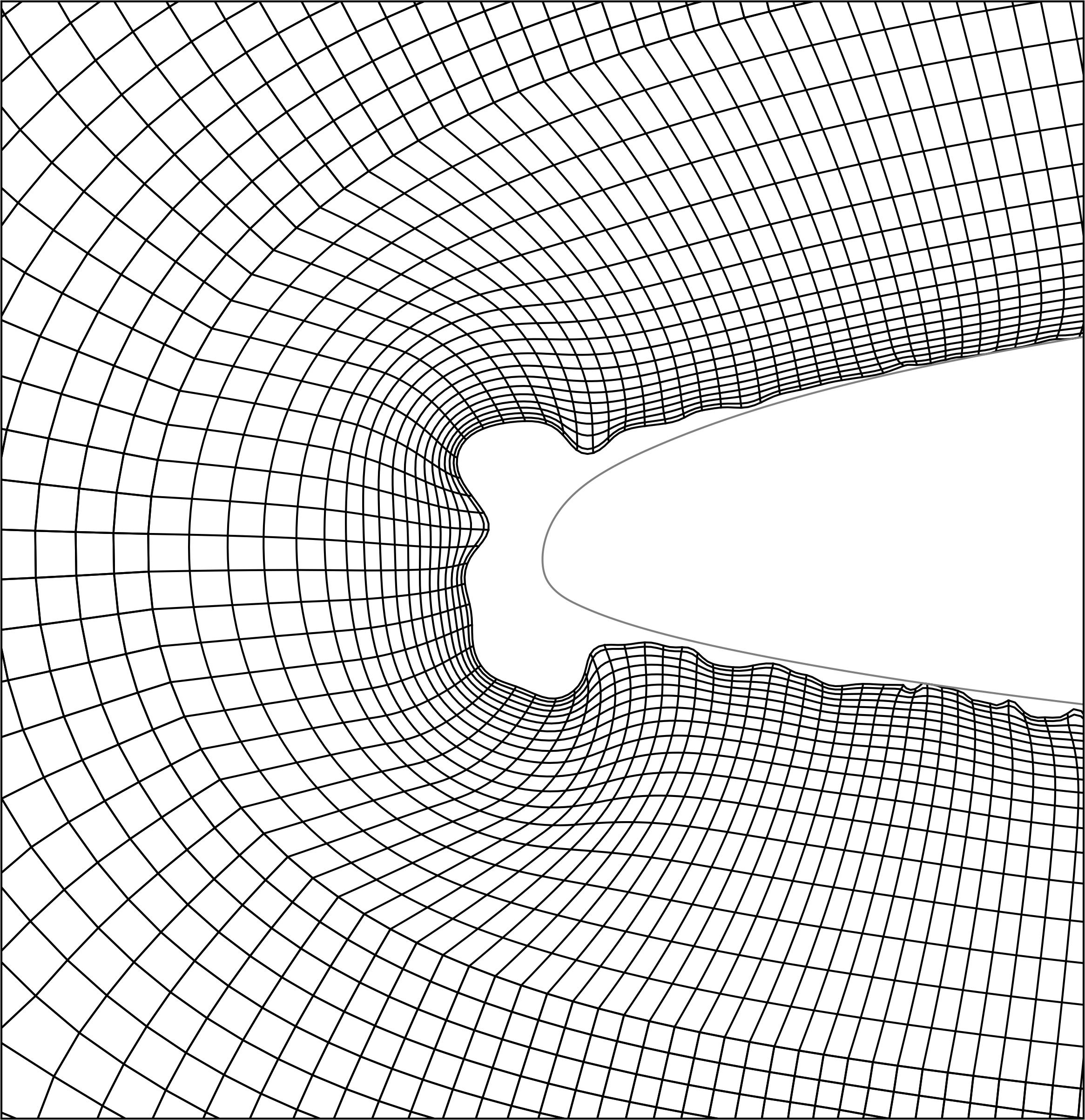}}
  
  \vspace*{5mm}
  \tikztitle{Hybrid C-grid layout}\\\vspace*{0.1cm}
  \fbox{\includegraphics[width=4.2cm, trim=2mm 2mm 2mm 2mm,clip=true]{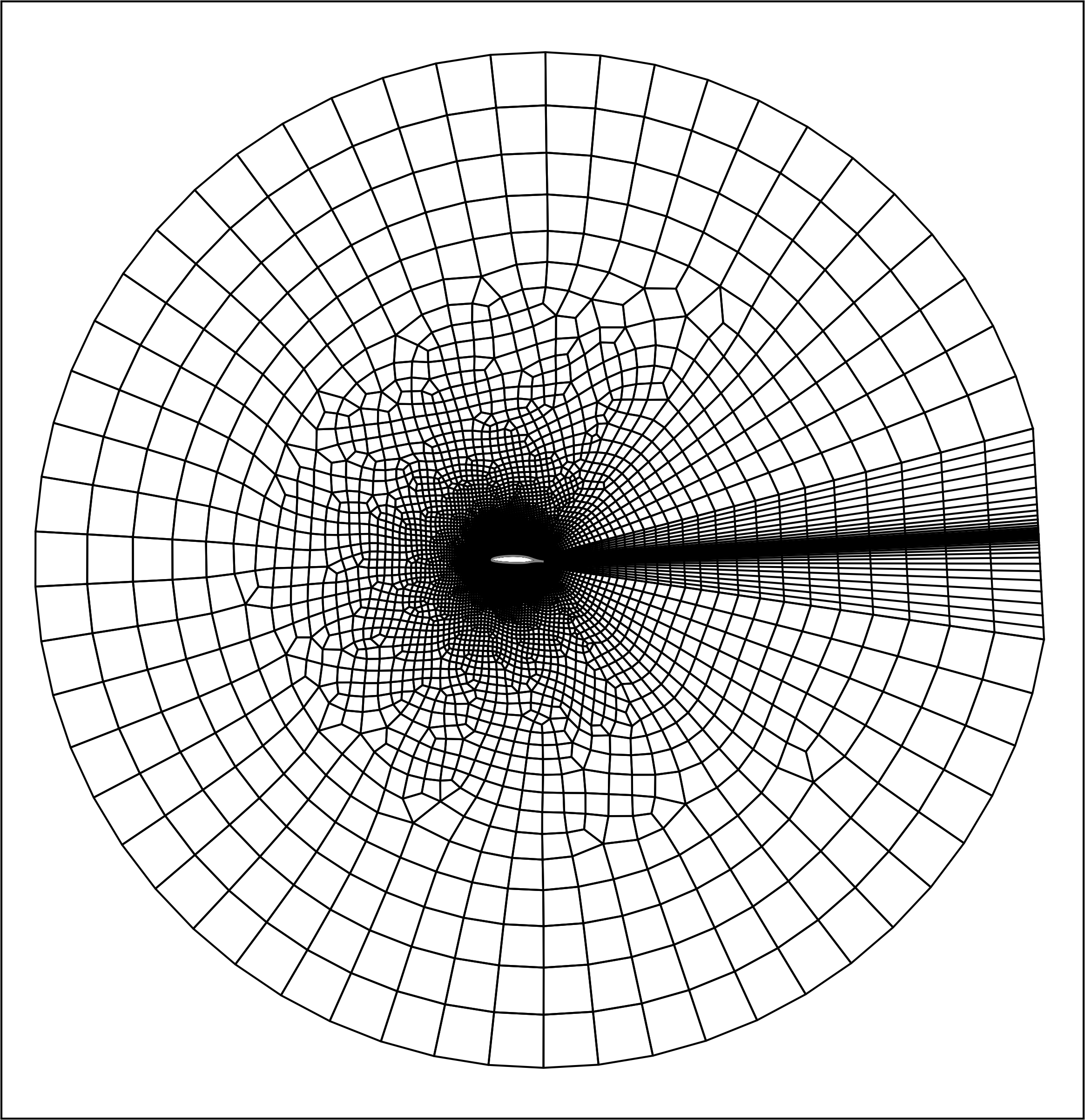}}\hspace{3mm}
  \fbox{\includegraphics[width=4.2cm, trim=2mm 2mm 2mm 2mm,clip=true]{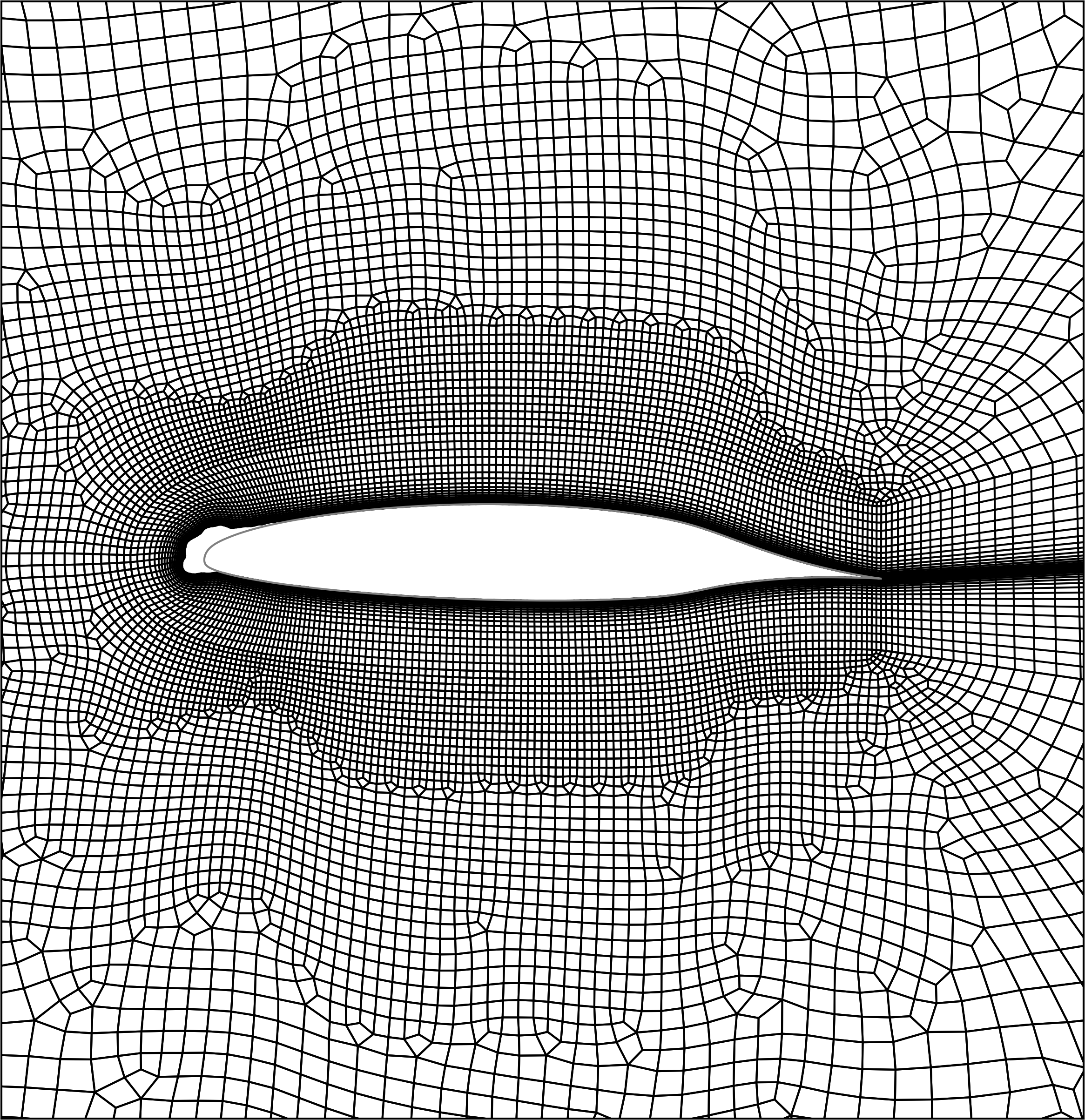}}
  \caption{Iced airfoil grid setup}
  \label{fig:icing_grid}
\end{figure}

For use with the flow solver FLEXI, polynomial curved DG elements are needed. The algorithm in Appendix \ref{sec:grid_appendix} is first carried out on a node-by-node basis, i.e. without regard for DG elements. Subsequently, patches of four by four nodes are grouped together and serve as geometric interpolation points of a curved DG element with a geometric polynomial degree of three.
The leading-edge regions of four grids created with the present algorithm are shown in \figref{fig:icing_grid} (a).

A structured wake grid is added to the boundary layer C-grid. The grid generation procedure is implemented in Python. A circular, unstructured far-field grid with non-curved elements is generated in Gmsh \cite{Geuzain2009Gmsh}. Both grids are transferred to an adapted version of the in-house open source\footnote{\texttt{https://github.com/hopr-framework/hopr}} high-order grid generator HOPR \cite{HOPR}, which merges the two parts, extrudes the two-dimensional grid in spanwise direction and writes a FLEXI-readable grid file. The result is shown in \figref{fig:icing_grid} (b).

\subsection{Baseline model setup and parameters}
\label{sec:cfd_prms}

The setup of the main simulation is designed to minimize the computational cost per sample, while still ensuring a high-quality LES. This allows to investigate UQ methods within a moderate computational budget. 

Simulations are carried out with a chord Reynolds number of $\reynolds_{\chord}=500,000$ and a Mach number of $\mach_{\infty}=0.4$, which roughly matches the conditions at the wing tip of a small wind turbine in strong winds. The angle of attack is $\aoa=3^{\circ}$. Simulations are run up to $\tEnd=10\chord/\velInf$. Time averaging of airfoil forces is started at $\tsym=5\chord/\velInf$, when a quasi-steady state is reached. This yields an averaging time span of $5\chord/\velInf$, during which time averages have converged sufficiently. For the DG method, a polynomial degree of $\ndg=5$ is used. The three-dimensional grid has a spanwise width of $0.05\chord$. 12 elements are used in spanwise direction. The radius of the far-field grid is $10\chord$. The near-wall resolution in wall units ensures $\dxp<40$, $\dyp<4.5$ and $\dzp<17$ in most regions of the airfoil. In streamwise and spanwise direction, the resolution is finer than typical requirements from the literature \cite{Piomelli2002,Georgiadis2010}. In wall-normal direction, the resolution exceeds recommended values of $\dyp<1$. It was found in previous applications that the polynomial solution approximation of the DGSEM entails that LES with a somewhat coarser resolution in wall-normal direction can maintain a high accuracy. However, wall-normal velocity gradients are very high around sharp ice horns, such that peak values up to $\dxp\approx85$, $\dyp\approx12$ and $\dzp\approx52$ occur in these very small and confined regions, which may nonetheless impair solution accuracy to some extent. 

Periodic boundary conditions are used in spanwise direction. The circular far-field boundary is divided into inflow- and outflow-semicircles based on the inflow direction. A constant Dirichlet boundary condition is used at the inflow. Characteristic outflow boundary conditions are used at the outflow. Sponge layers are used near the outflow boundaries to stabilize the outflow and reduce reflections. An adiabatic wall boundary condition is used at the iced airfoil with no thermodynamic distinction between iced and uniced regions.

An implicit no-model LES is carried out here, where the dissipation of the numerical scheme acts as a sub-grid-scale model. This approach has been validated and yields good results with the DGSEM \cite{Beck2014}. 

Three quantities of interest (QoIs) are considered: The time-averaged lift coefficient $\cl=\lift/(\chord\,\dzAF\,\dynpres)$, where $\lift$ is the airfoil lift force, $\chord$ is the airfoil chord length, $\dzAF$ is the spanwise width of the simulated airfoil segment and $\dynpres=1/2\,\dens_{\infty}\velInf^2$ is the free-stream dynamic pressure; the time-averaged drag coefficient $\cd=\lift/(\chord\,\dzAF\,\dynpres)$, where $\drag$ is the drag force; and the distribution of the time-averaged and spanwise-averaged local pressure coefficient $\cp=(\pres-\pres_{\infty})/\dynpres$ along the streamwise coordinate. 


\section{Uncertainty quantification (UQ)}
\label{sec:uq}

In numerical simulations, one or several quantities of interest (QoI) such as airfoil lift and drag are computed by a model evaluation. Normally, input parameters to the model are certain with a fixed value. In UQ, they are assumed to be random. More precisely, in this work it is assumed that the uncertain parameters can be expressed as a real-valued vector of $\ndimstoch$ independent random variables $\xiv = (\xicmp_1 , . . . , \xicmp_\ndimstoch )$ with image $\stochdom=\stochdom_{1}\times\dots\times\stochdom_{\ndimstoch}\subset\R^\ndimstoch$ defined on a probability space $(\samplespace, \sAlgebra, \pMeasure)$ and joint probability density function (pdf) $\pdf$. In this work, only uniformly distributed variables $\xiunif \sim \uniformdist(a,b)$ with lower and upper bounds $\unifLB$ and $\unifUB$ are used, but the described methods are not specific to these. 

Omitting the certain input parameters, the QoI can be expressed as a function of the random input $\scqoi(\xiv)$. It is also random. The goal of UQ is to obtain information about the distribution of the QoI, such as its mean $\E[\scqoi]$ and variance $\Var[\scqoi]$. The QoI can be real-valued, but also a vector or random field.

In this work, exclusively non-intrusive UQ methods are used. Their idea is to evaluate a standard model several times using different fixed values $\xiv_\iSample\in\stochdom,\,\,\iSample=1,\dots,\nSamples$ as input. 

The present study involves two steps: In the first step, a real-valued random vector representing the occurring ice shapes is obtained from a principal component analysis performed on an experimental data set. This random vector is then used as input to the second step, the UQ CFD simulations, which estimate the stochastic properties of the output quantities of interest (namely the forces on the airfoil) from this random input vector. 
This methodology was introduced by DeGennaro et al. \cite{DeGennaroData2015}. 

The rest of this Section first gives details on the data-integrated generation of the random vector. Subsequently, the used non-intrusive UQ methods are presented, namely NIPC method, followed by the two Monte Carlo methods MLMC and MFMC. Used UQ method parameters, the in-house code \pounce implementing all methods, and the used HPC resources are discussed at the end.


\subsection{Random input vector generation via principal component analysis (PCA)}
\label{sec:pca}

A set of experimentally measured ice shapes forms the basis of this study. An ideal data set would be quantitatively representative of the ice shapes occurring at an airfoil and of the frequency of their occurrence for typical aircraft flight envelopes or at wind turbine blades at a specific site. 
As an approximation to this, a set of 36 ice shapes from experiments in the NASA Icing Research Tunnel (IRT) at Glenn Research Center \cite{Addy2000} was used as simulation input. As the experiment, this study is based on the NASA/Langley NLF-0414F laminar flow airfoil \cite{Viken1987}.

\begin{figure}
  \centering
  \input{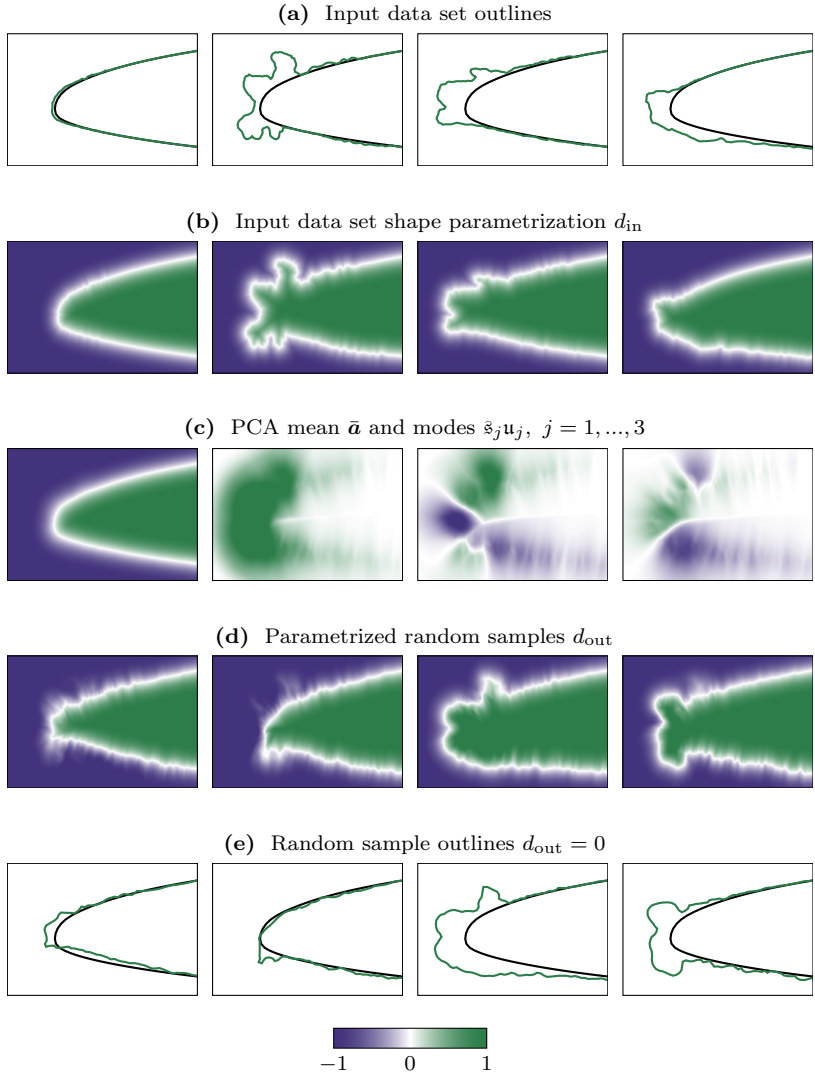}
  \caption[Ice shape parametrization and dimensionality reduction workflow.]{Ice shape parametrization and dimensionality reduction workflow. Only the first four samples or modes are shown for each step.}
  \label{fig:pca}
\end{figure}

Principal component analysis (PCA)\footnote{In \cite{DeGennaroData2015}, the term proper orthogonal decomposition (POD) is used instead of PCA for the same task. POD describes model order reduction of a deterministic physical model. PCA is the statistical term. The underlying algorithms are identical.} is used here to obtain a set of independent continuous random variables based on the (discrete) input data of experimental ice shapes. The implemented parametrization and PCA process is depicted in \figref{fig:pca} and open source\footnote{\texttt{https://github.com/flexi-framework/flexi-extensions/\allowbreak blob/\allowbreak pounce_ice/\allowbreak tools/\allowbreak generate_ice_shapes_and_mesh/get_pca_modes.py}}. 

As a starting points, four specimens from the input data set are shown in \figref{fig:pca} (a). A PCA generates a decomposition of a set of $\pcaNSamples$ input specimens (here the parametrized ice shapes), which are characterized by vectors of $\pcaNFeat$ features, which are denoted as $\pcaVecIn_{\pcaiIn}\in\R^{\pcaNFeat},\;\pcaiIn=1,...,\pcaNSamples$. This means that the input ice shapes must be parametrized, i.e. each shape must be converted into such a vector $\pcaVecIn_{\pcaiIn}$. The $i$th entry of each vector has to describe the same physical feature (here the same region of the ice shape) among specimens, which is why a set of points describing each shape outline are not a suitable choice. 
Therefore, instead a Cartesian grid is defined covering the iced region and a signed-distance function to the ice outline is evaluated at each grid point, and the values at all grid points are sorted into one vector $\pcaVecIn_{\pcaiIn}$ for each specimen. More precisely, this signed distance function is chosen as $\modeFunc=\arctan(3\min(\signedD/\dmax,1))$, where $\signedD$ is the distance to the nearest ice shape outline. The function is kept constant above a distance of $\dmax=0.05\,\chord$ and an arctangent scaling of the distance is used to reduce the influence of grid points further away from the ice shape outline. The resulting parametrized shapes are visualized in \figref{fig:pca} (b). We remark that in \cite{DeGennaroData2015}, only a binary parametrization $\modeFunc=\signSym$ with $\signSym=1$ inside the iced airfoil and $\signSym=-1$ outside is used for parametrization. In the present work, this approach led to very irregular and unphysical shapes and was therefore discarded.

The decomposition represents this discretized input data set in a coordinate system in which the input data is linearly uncorrelated. These basis vectors are later used as random input parameters to the simulation. First, the sample mean $\bar{\pcaVecIn}=1/\pcaNSamples\,\sum_{i=1}^{\pcaNSamples} \pcaVecIn_{i}$ is subtracted from the vectors $\pcaVecIn_{\pcaiIn}'=\pcaVecIn_{\pcaiIn}-\bar{\pcaVecIn}$. The centered vectors $\pcaVecIn_{\pcaiIn}'$ are assembled into a feature matrix $\pcaMatIn\in\R^{\pcaNFeat\times\pcaNSamples}$. The matrix is decomposed using a singular value decomposition (SVD), i.e. a factorization $\pcaMatIn=\pcaPC\pcaSVMat\pcaLS^{T}$ is calculated, where $\pcaPC\in\R^{\pcaNFeat\times\pcaNFeat}$ and $\pcaLS\in\R^{\pcaNSamples\times\pcaNSamples}$ are unitary matrices and $\pcaSVMat\in\R^{\pcaNFeat\times\pcaNSamples}$ is a rectangular diagonal matrix. The columns $\pcaPCcol_{\stochdimind}$ of $\pcaPC$ are then the principal components (i.e. the orthogonal basis vectors with unit length) and the singular values $\pcaSV_{\stochdimind}$ on the diagonal of $\pcaSVMat$ are the sample standard deviations in the direction of each principal component, sorted in descending order. 
The results of the SVD are shown in \figref{fig:pca} (c). The first plot shows the mean $\bar{\pcaVecIn}$. The other plots show the first three principal components scaled with their according singular values $\pcaSV_{\stochdimind}\pcaPCcol_{\stochdimind}$. Following the POD terminology, they are also called modes in the following.

The weighting of each of these scaled modes finally serves as a random input parameter to the simulation $\xicmp_{\stochdimind}$, which means that the shape is defined by setting $\pcaOut(\xiv)=\bar{\pcaVecIn}+\sum_{\stochdimind=1}^{\ndimstoch}\xicmp_{\stochdimind}\pcaSV_{\stochdimind}\pcaPCcol_{\stochdimind}$ and determining the random ice shape outline from the contour at $\pcaOut=0$. Four random samples of the function $\pcaOut$ are shown in 
\figref{fig:pca} (d) and the according shape outlines in \figref{fig:pca} (e).
Due to the scaling of the modes with the singular values, the modes $\pcaSV_{\stochdimind}\pcaPCcol_{\stochdimind}$ have unit sample variance in the original data set.

PCA generally does not provide any information on the probability distribution of the principal components apart from their mean and variance, so assumptions have to be made. Usually, a normal distribution is used. In this work, a uniform distribution is chosen instead, as the boundedness of the signed distance function serving as input to the PCA motivates the choice of a bounded distribution. 
Apart from this assumption on the random distribution, there are further potential sources of error: In PCA, the resulting random variables are linearly uncorrelated in the input data set, but not necessarily independent, i.e. non-linear dependencies can still exist. Moreover, the limited size of the input data set size of 36 specimens can lead to inaccurate estimates.

These errors can lead to some unphysical phenomena in the sample ice shapes: The generated outlines sometimes lie inside the original airfoil, as can be seen in \figref{fig:pca} (e). In rare cases, islands of ice can form in the air which are not connected to the rest of the ice or the airfoil. As a remedy, the outline is replaced by the airfoil contour wherever it lies inside the airfoil. Unconnected islands of ice are simply omitted.  
Moreover, the distribution of each random variable is set to $\xicmp_{\stochdimind}\sim\uniformdist(-0.75,0.75)$. This distribution only has a standard deviation of approximately 0.43, which is less than the normalized standard deviation of 1.0 in the input data set, such that only a subset of cases near the input data average is considered. This excludes most of the unphysical shapes which are generated due to the linearity of the PCA. 

While these adjustments impose limitations on the quantitative representation of the input data in the generated random vector, the data-driven approach is still a big improvement over purely heuristic assumptions on the random input found in many UQ studies. 

PCA is a dimensionality reduction technique. The first $n$ principal components form the basis in $\R^{n}$ which optimally explains the variance of the input data set. To reduce the stochastic dimensionality of the problem, optionally only the first few random variables can therefore be considered, while the rest is omitted. This reduction will be exploited later for stochastic dimensionality reduction in NIPC.


\subsection{Non-intrusive UQ methods}

The three non-intrusive UQ methods used in this study to propagate the uncertainty through the CFD model are presented in the following.

\subsubsection{Non-intrusive polynomial chaos (NIPC)}
\label{sec:nipc}

In non-intrusive polynomial chaos, the quantity of interest is expanded into an infinite polynomial series
\begin{equation}\label{eq:polydef}
  \scqoi(\xiv) = \sum_{\stochind\in\indspace} \coef{\scqoi}_{\stochind}\basis_\stochind(\xiv)
\end{equation}
with $\indspace=\N_0^\ndimstoch$, the polynomial coefficients $\coef{\scqoi}_\stochind  \in \R$ and the basis functions  $\basis_\stochind(\xiv)$, which are products of univariate polynomials 
\begin{equation}
  \basis_\stochind(\xiv) = \prod_{\stochdimind=1}^{\ndimstoch} \basis_{\stochind_\stochdimind}(\xicmp_\stochdimind).
\end{equation}
with the multi-index $\stochind = (\stochind_1,...,\stochind_\ndimstoch)^T \in \N_0^\ndimstoch$. As a univariate basis, we choose basis vectors from the Askey scheme, which are orthonormal with respect to the probability distribution of the random input vector, such that
\begin{equation}
  \langle\basis_\stochind,\basis_{\iota}\rangle = \prod_{\stochdimind=1}^{\ndimstoch} \delta_{\stochind_{\stochdimind},\iota_{\stochdimind}} = \delta_{\stochind,\iota}, 
\end{equation}
with the Kronecker delta $\delta_{\stochind,\iota}$ and the inner product defined as
\begin{equation} \label{eq:innerprod}
  \langle \varg(\xiv),\varh(\xiv)\rangle  \coloneqq \int_{\xiv\in\stochdom} \varg(\xiv)\varh(\xiv)\pdf ~\mathrm{d\xiv}.
\end{equation}
Here, we truncate the series with an isotropic tensor product ansatz
\begin{equation} \label{eq:idxset_tp}
\indspace \coloneqq \Big\{ \stochind \in \N_0^\ndimstoch \;\Big|\; \stochind_\stochdimind \le \maxstochpolydeg,\;\; \stochdimind=1,...,\ndimstoch \Big\}.
\end{equation}
We retrieve the polynomial coefficients $\coef{\scqoi}_\stochind$ from a Galerkin projection in the stochastic space 
\begin{equation}\label{eq:projection}
  \coef{\scqoi}_\stochind =  \langle \scqoi,\basis_\stochind \rangle. 
\end{equation}
To evaluate the stochastic integral in \equref{eq:projection}, we use a tensor product of Gaussian quadrature rules 
\begin{equation}\label{eq:projection_nisp}
  \coef{\scqoi}_\stochind =  \langle \scqoi,\basis_{\stochind} \rangle \approx 
  \sum_{\xiQuadIndex=1}^{\nQP^{\ndimstoch}} \sdqoi(\xiv_{\xiQuadIndex})\,\basis_{\stochind}(\xiv_{\xiQuadIndex})\,\quadRweight{\xiQuadIndex}
\end{equation}
with the quadrature weights $\quadRweight{\xiQuadIndex}$ and the multi-index set
\begin{equation}
\xiQuadIndex \in \Pidxset:=
\{\xiQuadIndex=(\xiQuadIndex_1,\ldots,\xiQuadIndex_\ndimstoch)^\top \in\mathbb{N}^\ndimstoch
~\big|~\xiQuadIndex_{\stochdimind}\leq\nQP, ~\stochdimind=1,\ldots,\ndimstoch\}.
\end{equation}
using $\nQP=\maxstochpolydeg+1$ quadrature points in each dimension. Mean and variance are then approximated as 
\begin{equation} \label{eq:mean_and_var}
  \E(\scqoi) \approx \coef{\scqoi}_0 \quad\text{and}\quad \Var(\scqoi) \approx \sum_{\stochind \in \indspace\backslash (0,...,0)^T} \coef{\scqoi}_\stochind^2, 
\end{equation}
following their definition and the orthonormality of the basis. 

NIPC is efficient for smooth QoIs where the interpolation is a good approximation, and for few uncertain variables, as the required cost grows exponentially with the number of stochastic dimensions.

\subsubsection{Monte Carlo (MC)}

In Monte Carlo methods, random samples are drawn from the random input vector and the computational model is evaluated for each sample point. Mean $\E[\scqoi]$ and variance $\Var[\scqoi]$ of the QoI are then gained from the unbiased estimators
\begin{equation}\label{eq:mc_mean}
  \E[\scqoi]\approx\dE_\nSamples[\sdqoi] = \frac{1}{\nSamples} \sum_{\iSample=1}^{\nSamples} \sdqoi_\iSample
\end{equation}
and
\begin{equation}\label{eq:mc_variance}
  \Var[\scqoi]\approx\dVar_\nSamples[\sdqoi] 
  = \frac{1}{\nSamples-1}  \sum_{\iSample=1}^{\nSamples} 
  \left(\sdqoi_\iSample - \dE_\nSamples[\sdqoi] \right)^2.
\end{equation}
The mean squared error (MSE) of the MC mean estimator is 
\begin{equation}\label{eq:mc_stocherrdef}
\err^2 \coloneqq \E\big[(\dE_\nSamples[\sdqoi]-\E[\sdqoi])^2\big]  = \frac{1}{\nSamples} \Var[\sdqoi]. 
\end{equation}
MC convergence is very robust and independent of the number of uncertain parameters. However, the estimator error shows only half-order convergence, such that many evaluations of the baseline model are needed for accurate results. This motivates using error reduction techniques.

\subsubsection{Multilevel Monte Carlo (MLMC)}
\label{sec:mlmc}

In the MLMC method developed by Heinrich \cite{Heinrich2001} and Giles \cite{Giles2008}, few sample simulations with a high numerical accuracy and many with computationally cheaper, less accurate models are combined to achieve both overall high numerical and stochastic accuracy at moderate cost. In addition to the baseline model, a series of models with coarser numerical resolutions and an otherwise identical setup are defined, which results in the set of models $\{\sdqoi^\levelind(\xiv) \}_{ \levelind = 1, ..., \nlevels}$, where the resolution increases with $\levelind$. In this work, we achieve this change in resolution not by a change of the grid spacing, but a change of the polynomial degree $\ndg$, which is introduced as multi-order Monte Carlo in \cite{Motamed2018}. 

To derive the estimator, the finest resolved (baseline) model $\sdqoi^{\nlevels}$ is expressed as the telescopic sum 
\begin{equation}
  \sdqoi^{\nlevels} = \sdqoi^{1} + \sum_{\levelind=2}^{\nlevels} \sdqoi^{\levelind}-\sdqoi^{\levelind-1} = \levelsum \Delta \sdqoi^{\levelind}
\end{equation}
with the differences between resolution levels
$\Delta \sdqoi^{\levelind} \coloneqq \sdqoi^{\levelind}-\sdqoi^{\levelind-1}$
and the auxiliary $\sdqoi^{0}\coloneqq 0$. For the expectation $\E[\scqoi]$, linearity yields 
\begin{equation}
  \E[\scqoi]\approx\E[\sdqoi^{\nlevels}] = \levelsum \E[\Delta \sdqoi^{\levelind}].
\end{equation}
The expectations $\E[\Delta \sdqoi^{\levelind}]$ are now replaced by Monte Carlo estimators with different numbers of samples $\nSamples^{\levelind}$ to obtain the MLMC estimator
\begin{equation} \label{eq:mlmc_exp}
  \E[\scqoi]\approx\E[\sdqoi^{\nlevels}]\approx\dEMLMC[\sdqoi^{\nlevels}] 
  = \levelsum \dE_{\nSamples^{\levelind}}[\Delta \sdqoi^{\levelind}]
  = \levelsum \frac{1}{\nSamples^{\levelind}} \sum_{\iSample^{\levelind}=1}^{\nSamples^{\levelind}} \left(\sdqoi_{\iSample^{\levelind}}^{\levelind} - \sdqoi_{\iSample^{\levelind}}^{\levelind-1}\right)
\end{equation}
with the samples $\sdqoi_{\iSample^{\levelind}}^{\levelind}=\sdqoi^{\levelind}(\xiv_{\iSample^{\levelind}})$.
Each level difference $\left(\sdqoi_{\iSample^{\levelind}}^{\levelind} - \sdqoi_{\iSample^{\levelind}}^{\levelind-1}\right)$ builds on computations of the same sample with different resolution models. However, the index $\iSample^{\levelind}$ indicates that samples are drawn independently for each level difference mean estimator, i.e. the difference between Levels 1 and 2 is estimated with different realizations of the random vector than the difference between Levels 2 and 3.

Apart from the mean, the variance of the QoI $\scqoi$ is also of interest. Here, the estimator 
\begin{equation} \label{eq:mlmc_var}
  \Var[\scqoi]\approx\Var[\sdqoi]\approx\dVarMLMC[\sdqoi^{\nlevels}] 
  = \levelsum \left( \dVar_{\nSamples^{\levelind}}[\sdqoi^{\levelind}] - 
  \dVar_{\nSamples^{\levelind}}[\sdqoi^{\levelind-1}] \right)
\end{equation}
is used, with the variance estimator $\dVar_{\nSamples}[\cdot]$ defined in \eqref{eq:mc_variance}. It is an unbiased estimator of $\Var[\sdqoi]$. It was presented in \cite{Bierig2015}, where error bounds are discussed. 

The MLMC estimator error is 
\begin{equation}\label{eq:mlmc_totalerr}
  \err_{\MLMCop}^2 = \levelsum \frac{1}{\nSamples^{\levelind}} \Var[\Delta\sdqoi^{\levelind}]
\end{equation}
The MLMC error compared with the error of the standard MC method with a given computational budget depends on the convergence of the numerical method, which determines the variance of the level differences $\Var[\Delta\sdqoi^{\levelind}]$, and on the computational cost of a sample evaluation on each level, which determines the affordable number of evaluations $\nSamples^{\levelind}$. 

The number of samples on every level which yields the lowest error can be obtained from an optimization problem with a computational budget as a constraint (alternatively, computational cost can be minimized in a similar fashion for a given MSE). Again, driving factors are the convergence of the computational method and sample evaluation costs. 

Those quantities are normally unknown a priori, but can be estimated from some pilot samples. This two-iteration approach (compute $\mathcal{O}(10)$ pilot samples, estimate optimal sample numbers, compute rest of the samples) is state of the art \cite{PyMLMCSukys2017}. However, it is desirable to base the sample number estimates on the highest possible number of samples to avoid sub-optimal sample distributions among levels. To this end, a method presented in \cite{Beck2020} is used to update these estimates in a three or four iteration approach. A larger number of iterations creates scheduling drawbacks, as it is advantageous to group large numbers of samples into batches without intermittent post-processing. 

For field-valued QoIs, the integration approach presented in \cite{Muller2013} is used. 

\subsubsection{Multifidelity Monte Carlo (MFMC)}
\label{sec:mfmc}

In MFMC, in addition to the baseline (high fidelity, or short HF) model $\sdqoi^{\HF}\coloneqq\sdqoi^{1}$, a series of low fidelity (LF) models $\sdqoi^{2},...,\sdqoi^{\nModels}$ is defined which are less computationally expensive. These LF models can be arbitrary as long as their QoI is correlated with that of the HF model. So in addition to coarse grid resolutions found in MLMC, simpler physical models, lower dimensional models or surrogate models are possible.

Here, an approach based on control variates (CV) presented in \cite{Peherstorfer2016} is used. CV is a variance reduction technique where a correlated variable (here the output of a low fidelity model, the control variate) with a known mean is used to reduce the MC estimator error. To this end, the difference between the same MC estimator and the known mean of the CV is added to the original MC estimator. The mean of low fidelity models is unknown, but it can be estimated more accurately than that of the HF model, since the model evaluation is less costly. Several low fidelity models with increasingly lower cost can be used jointly as a sum of control variates. This yields the MFMC estimator 
\begin{equation} \label{eq:mfmc_exp}
  \arraycolsep=1pt\def\arraystretch{2.5}
  \begin{array}{ccccccccl}
  \dEMFMC[\sdqoi^{1}]
    &\ds = &\ds \dE_{\nSamples^{1}}[\sdqoi^{1}]
    &\ds + &\ds \modelsum{2} \mfmcAl^{\iModel}\Big(
           &\ds \dE_{\nSamples^{\iModel}}[\sdqoi^{\iModel}]
         &\ds - &\ds \dE_{\nSamples^{\iModel-1}}[\sdqoi^{\iModel}] &\ds \Big)\\
    &\ds = &\ds \dfrac{1}{\nSamples^{1}} \sum_{\iSample=1}^{\nSamples^{1}} \sdqoi^{1}(\xiv_{\iSample})
    &\ds + &\ds \modelsum{2} \mfmcAl^{\iModel}\bigg(
        &\ds \dfrac{1}{\nSamples^{\iModel}} \sum_{\iSample=1}^{\nSamples^{\iModel}} \sdqoi^{\iModel}(\xiv_{\iSample})
      &\ds - &\ds \dfrac{1}{\nSamples^{\iModel-1}} \sum_{\iSample=1}^{\nSamples^{\iModel-1}} \sdqoi^{\iModel}(\xiv_{\iSample}) &\ds \bigg),
\end{array}
\end{equation}
where $\mfmcAl^{\iModel}\in\R$ are the CV coefficients. Note that the models are ordered by the number of sample evaluations $\nSamples^{\iModel}<\nSamples^{\iModel-1}$ for all $\iModel$. Furthermore, in contrast to MLMC, the samples used for the estimator $\dE_{\nSamples^{\iModel}}[\cdot]$ are always a subset of the samples used for the estimator $\dE_{\nSamples^{\iModel-1}}[\cdot]$.
An optimization problem is solved in \cite{Peherstorfer2016} minimizing the estimator error for a given computational budget to find the optimal sample numbers $\nSamples^{\iModel}$, the optimal CV coefficients $\mfmcAl^{\iModel}$, but also the optimal choice of low fidelity models: A set of models is proposed to an algorithm which finds the subset of models promising the highest efficiency and discards the other models. For the proof, algorithms and formulae for $\nSamples^{\iModel}$ and $\mfmcAl^{\iModel}$, the reader is referred to \cite{Peherstorfer2016}. The mean squared estimator error resulting from the minimization problem is given by
\begin{equation}\label{eq:mfmc_mse2}
  \err_{\MFMCop}^2 = 
  \frac{\Var[\sdqoi^{1}]\work^{1}}{\cBudget}
  \left( \modelsum{1} \sqrt{\frac{\work^{\iModel}}{\work^{1}}
  \Big((\pears^{1,\iModel})^2-(\pears^{1,\iModel+1})^2\Big)}\right)^2,
\end{equation}
with the computational budget $\cBudget$, the computational cost for a model evaluation $\work^{\iModel}$, and the Pearson correlation coefficient between the high fidelity model and model $\iModel$
\begin{equation} \label{eq:pearson_coeff}
  \pears^{1,\iModel}=\frac{\E[(\sdqoi^{1}-\E[\sdqoi^{1}])(\sdqoi^{\iModel}-\E[\sdqoi^{\iModel}])]}
                          {\sqrt{\Var[\sdqoi^{1}]\Var[\sdqoi^{\iModel}]}}.
\end{equation}
In \eqref{eq:mfmc_mse2}, only the optimal subset of low fidelity models is considered. The factor in front of the sum equals the variance of the ordinary Monte Carlo estimator using only the high-fidelity model and the same computational budget and the squared sum equals the variance reduction.

As in MLMC, model cost, variances and correlation coefficients are unknown a priori, but have to be estimated from some pilot samples.

The MFMC publication \cite{Peherstorfer2016} suggests not to use the samples used for these estimates for the eventual MFMC estimator, but to use independently drawn new samples instead. This is necessary to ensure unbiasedness of the estimator, but incurs significant additional cost. In practical applications, mostly no difference in accuracy is observed when re-using pilot samples for the eventual estimator. In this work, pilot samples for the estimation of model cost, variances and correlation coefficients are therefore re-used for the MFMC estimator to achieve a higher efficiency. Moreover, the estimates for the optimal coefficients $\mfmcAl^{\iModel}$ are updated after computing all samples, such that they are based on a larger sample size, which promises accuracy gains.

MFMC optimal sample number estimates are based on the correlation estimates of each low fidelity model to the highest fidelity model (in contrast to comparing two different low-resolution models in MLMC), so the sample size is constrained to the number of samples with the high fidelity model, which is usually in a similar range $\mathcal{O}(10)$ as the number of pilot samples. The expected gain of a multi-iteration approach as described and referenced above for MLMC is therefore low. Only two iterations are therefore used, as proposed in \cite{Peherstorfer2016}.

\subsubsection{UQ method parameters}
\label{sec:uq_prms}

For the NIPC simulation, only the first two PCA modes are considered. A stochastic polynomial degree of $\maxstochpolydeg=4$ is used with a tensor-product quadrature with $5\times5=25$ points resulting in a total of 25 large-eddy simulations. 

For the MLMC and MFMC simulations, all 35 uncertain modes are considered, since Monte Carlo convergence is not dimension dependent, such that more modes promise higher accuracy at a similar cost. A computational budget of one million CPUh is prescribed for both simulations. 10 pilot samples are used for each model and level. The sample number and choice of models is optimized with respect to the pressure coefficient $\cp$ as a QoI. This leads to sub-optimal choices for the other QoIs. 
For MLMC, coarse levels are realized by varying the DG polynomial degree to $\ndg=1$ and $\ndg=3$, all on the same grid. For MFMC, low-fidelity models are also realized by varying the DG polynomial degree to $\ndg=3$ and $\ndg=1$ on the same grid. Additionally, two-dimensional simulations are used as low-fidelity models with $\ndg=5$, $\ndg=3$ and $\ndg=1$ and an otherwise identical setup. Note that an implicit LES closure is used, such that the flow in the two-dimensional simulations stays laminar. This may impair accuracy of these low-fidelity models, but if the models are correlated to the high-fidelity model with respect to the effect of the ice shape, they can still increase the overall efficiency due to their low cost. 
For the MLMC simulation, four iterations are used, as described in \secref{sec:mlmc}. In the MFMC simulation, the pilot samples are reused and the coefficients $\mfmcAl^{\iModel}$ are updated after the second iteration as suggested in \secref{sec:mfmc}.


\subsection{Implementation}
\subsubsection{UQ management framework: \pounce}
\label{sec:pounce}

The above methods are implemented in the in-house UQ management framework \pounce, which is open source\footnote{\texttt{https://github.com/JakobBD/pounce}} and published in \cite{Duerrwaechter2023}. 
Its purpose is to draw, pre-process, run, post-process and evaluate samples. In MLMC and MFMC methods, it runs in several iterations, as sample numbers are determined adaptively. It follows a novel scheduling strategy for fully automated and efficient runs on HPC clusters. All samples of one model and one iteration are grouped to a sample batch with common file I/O and one scheduler job per batch. For this, interaction with the cluster scheduler is also automated. The source code is written in Python and modular, such that new functionality (new UQ methods, new clusters and new baseline solvers) can be added easily. Further details can be found in \cite{Duerrwaechter2023} and in the code documentation in the above repository.

\subsubsection{Sample simulations}
\label{sec:hpc}

UQ simulations were run on the high-performance computing cluster `Hawk' at the High-Performance Computing Center Stuttgart (HLRS) using \pounce. Sample simulations were carried out with a FLEXI version adapted for use in combination with the UQ framework \pounce. 

The Python framework as well as pre- and post-processing were run on one node with up to 128 cores. The main simulation sample batches were run on up to 512 nodes (65,536 cores) in each of the three methods. The uniced simulation took approximately 20,000 CPUh. The overall computational cost for the NIPC simulation was 544,000 CPUh. This is 22,000 CPUh per sample, slightly more than in the uniced case, due to a smaller time step on the distorted grid.  
In the Monte Carlo methods, iced sample simulation with the finest resolution $\ndg=5$ took on average 40,000 CPUh.
Again, the increased computational cost is due to a smaller timestep following a distorted grid. 
The effect is much stronger here than in NIPC. The first two PCA modes used in NIPC have a rather low spatial frequency and thus do not cause sharp curves in the airfoil outline, such that the near-wall distortion is less severe and the time step is closer to the uniced case.  

For the MLMC simulations, $\vec{\nSamples}=(379,\,57,\,15)^T$ samples were computed on the three levels. Details are given in \secref{sec:mlmc_method_results}. The computational budget was kept with minor deviations due to slightly sub-optimal machine use and inaccuracies in mean sample cost estimation. For MFMC, the optimal set of models was determined to be the high-fidelity model along with the three two-dimensional models, while the two three-dimensional low-fidelity models were discarded after the pilot sample computation. A total of $\vec{\nSamples}=(23,\,10,\,10,\,174,\,587,\,3676)^T$ samples were computed with each model. Details are again given in \secref{sec:mfmc_method_results}. The computational cost for the pilot samples of the discarded models is not counted towards the computational budget.

\section{Results}
\label{sec:results}

UQ simulation results are presented in the following. In \secref{sec:cp_curves}, mean and standard deviation of the pressure coefficient curves are discussed. Mean, standard deviation and estimated error are compared across methods for each QoI in \secref{sec:mean_std_table}. In Sections \ref{sec:mlmc_method_results} and \ref{sec:mfmc_method_results}, MLMC and MFMC methods are evaluated in more detail by discussing method parameters which resulted from the optimization procedures inherent in the methods. Lastly, in \secref{sec:flowfields_results}, the fluid dynamics underlying the results are interpreted in more depth by assessing flow fields of sample simulations and NIPC response surfaces of the lift and drag coefficients. 

\subsection{Pressure coefficient curves}
\label{sec:cp_curves}

\begin{figure}[t!]
  \centering
  \begin{tikzpicture}

\def\scaleSigma{1.0}

\pgfplotstableread{plots/icing/csv_cp/sc_ps.csv}{\scpstable}
\pgfplotstableread{plots/icing/csv_cp/sc_ss.csv}{\scsstable}
\pgfplotstableread{plots/icing/csv_cp/mlmc_ps.csv}{\mlmcpstable}
\pgfplotstableread{plots/icing/csv_cp/mlmc_ss.csv}{\mlmcsstable}
\pgfplotstableread{plots/icing/csv_cp/mfmc_ps.csv}{\mfmcpstable}
\pgfplotstableread{plots/icing/csv_cp/mfmc_ss.csv}{\mfmcsstable}
\pgfplotstableread{plots/icing/csv_cp/noice.csv}{\noicetable}
\pgfplotstableread{plots/icing/csv_cp/xy.csv}{\xytable}

\begin{groupplot}[group style={group size=1 by 3, vertical sep = -0.5cm},
width=0.85 \linewidth,
height=0.4 \linewidth,
scale only axis,
]

\nextgroupplot[
title= \tikztitle{Mean},
scale only axis,
xmode=linear,
xminorticks=true,
ymode=linear,
xmin=-0.10,
xmax=1.001,
yminorticks=true,
no markers,
xlabel={$x/\chord$},
ylabel={$\dE[\cp]$},
y dir = reverse,
legend columns=1, 
legend pos = north east,
legend style={font= \scriptsize},
legend cell align={left},
]
\addplot[thick,black,dashed] table [x=x , y=mean] {\noicetable};
\addlegendentry{uniced}
\addplot[very thick,v3_1] table [x=x , y=mean] {\scpstable};
\addlegendentry{NIPC}
\addplot[very thick,v3_2] table [x=x , y=mean] {\mlmcpstable};
\addlegendentry{MLMC}
\addplot[very thick,v3_3] table [x=x , y=mean] {\mfmcpstable};
\addlegendentry{MFMC}

\addplot[very thick,v3_1, forget plot] table [x=x , y=mean] {\scsstable};
\addplot[very thick,v3_2, forget plot] table [x=x , y=mean] {\mlmcsstable};
\addplot[very thick,v3_3, forget plot] table [x=x , y=mean] {\mfmcsstable};

\addplot[name path=mfmcpsLower, draw=none, forget plot]  table [x=x, y expr=\thisrow{mean}-\scaleSigma*\thisrow{stddev}] {\mfmcpstable};
\addplot[name path=mfmcpsUpper, draw=none, forget plot]  table [x=x, y expr=\thisrow{mean}+\scaleSigma*\thisrow{stddev}] {\mfmcpstable};
\addplot[stdgray,fill opacity=1.0] fill between[of=mfmcpsLower and mfmcpsUpper];
  \addlegendentry{$\expec\pm\stddev$}
\addplot[name path=mfmcssLower, draw=none]  table [x=x, y expr=\thisrow{mean}-\scaleSigma*\thisrow{stddev}] {\mfmcsstable};
\addplot[name path=mfmcssUpper, draw=none]  table [x=x, y expr=\thisrow{mean}+\scaleSigma*\thisrow{stddev}] {\mfmcsstable};
\addplot[fill opacity=0.2] fill between[of=mfmcssLower and mfmcssUpper];

\nextgroupplot[
scale only axis,
axis equal,
xmode=linear,
ymode=linear,
xmin=-0.10,
xmax=1.001,
no markers,
ticks = none,
axis line style={draw=none}
]
\addplot[black] table{\xytable};

\nextgroupplot[
title= \tikztitle{Standard deviation},
scale only axis,
xmode=linear,
xminorticks=true,
ymode=linear,
xmin=-0.10,
xmax=1.001,
yminorticks=true,
ytick = {-0.4,-0.2,0,0.2,0.4,0.6,0.8},
yticklabels = {0.4,0.2,0,0.2,0.4,0.6,0.8},
no markers,
xlabel={$x/\chord$},
ylabel={$\dSD[\cp]$},
legend columns=1, 
legend pos = north east,
legend style={font= \scriptsize},
legend cell align={left},
]
\addplot[very thick,v3_1] table [x=x , y expr=\thisrow{stddev}+0.0] {\scsstable};
\addlegendentry{NIPC}
\addplot[very thick,v3_2] table [x=x , y expr=\thisrow{stddev}+0.0] {\mlmcsstable};
\addlegendentry{MLMC}
\addplot[very thick,v3_3] table [x=x , y expr=\thisrow{stddev}+0.0] {\mfmcsstable};
\addlegendentry{MFMC}

\draw ({axis cs: 0,0}-|{rel axis cs:0,0})--({axis cs:0,0}-|{rel axis cs:1,0});

\addplot[very thick,v3_1] table [x=x , y expr=-\thisrow{stddev}+0.0] {\scpstable};
\addplot[very thick,v3_2] table [x=x , y expr=-\thisrow{stddev}+0.0] {\mlmcpstable};
\addplot[very thick,v3_3] table [x=x , y expr=-\thisrow{stddev}+0.0] {\mfmcpstable};

\end{groupplot}

\end{tikzpicture}%
  \caption[Mean and standard deviation of pressure coefficient $\cp$.]{Mean and standard deviation of pressure coefficient $\cp$. The gray shading is the standard deviation of the MFMC simulation. The standard deviation axis of the pressure side in the bottom plot is flipped downwards.}
  \label{fig:mfmc_cp_expstd}
\end{figure}
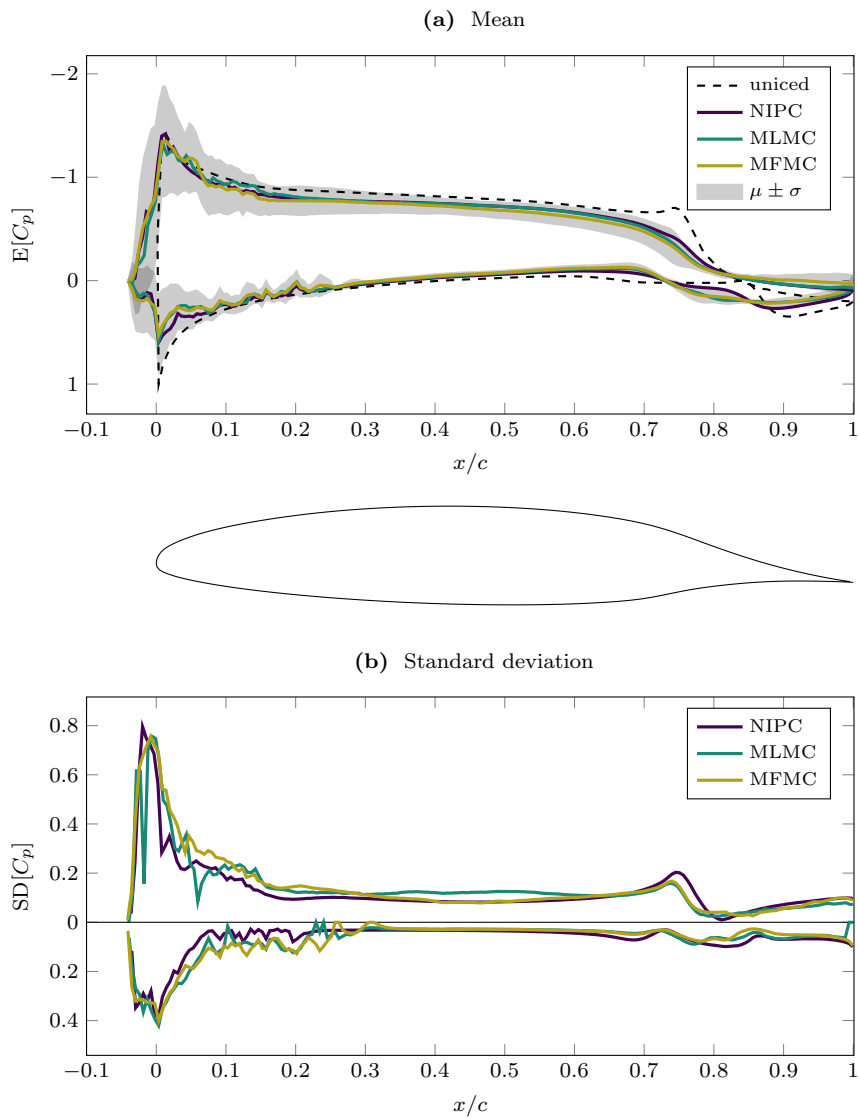

\figref{fig:mfmc_cp_expstd} shows mean (top) and standard deviation (bottom) of the pressure coefficient $\cp$ over the streamwise coordinate. The standard deviation of the MFMC simulation is shown in the mean plot as a gray shaded area. The standard deviation of the pressure side is flipped below the plot axis for a better distinction of the curves. 

The mean curves show notable differences to the uniced case. Non-zero values are present even for $\xSym<0$, as the ice extends forward from the airfoil leading edge. The iced regions near the leading edge are governed by irregularities and peaks on both suction side and pressure side. On the pressure side, the uncertain position of the stagnation point appears as a reduced amplitude of the according peak. The mean suction peak has almost the same height as that of the uniced case, as the effects of lift-reducing flow separation and locally high velocities cancel out in the mean. The curves of the uniced airfoil, after a region of relatively constant pressure near the center of the airfoil, exhibit a bump followed by increasing pressure on both sides (the ordinate is inverted in $\cp$-plots), which is caused by laminar separation bubbles with laminar-to-turbulent transition.
The iced mean curves show similar patterns. However, the pressure level is on average higher on the suction side due to separated flow caused by ice horns on the suction side in some cases. The increasing pressure sets in earlier compared with the uniced cases. The likely reason are disturbances in the flow due to the leading edge ice, which leads to earlier transition and thus altered or removed  separation bubbles.

Agreement between the Monte Carlo methods is excellent for the mean. The NIPC method shows very similar results in the forward half of the airfoil, but the adverse pressure gradient in the aft region is located further towards the trailing edge (especially on the pressure side), which indicates a later laminar-to-turbulent transition. This effect is also likely due to the more regular ice shapes and the less disturbed flow as a consequence of it.

The standard deviation is largest near the leading edge, and larger on the suction side than on the pressure side. Agreement between the methods is good. The NIPC method shows a slightly lower standard deviation in the front region, which is also an effect of the missing high-frequency modes. Uncertainties in the position of the adverse pressure gradient lead to a small peak of standard deviation on the suction side at $\xSym/\chord\approx0.75$, which is largest for the NIPC method.

\subsection{Scalar mean, standard deviation and error values}
\label{sec:mean_std_table}

\tabref{tab:icing_results} lists the estimated mean, standard deviation (StD) and root mean squared error (RMSE) for the three investigated QoIs and the three investigated methods along with the uniced case. Error estimates are available as $\err_{\MLMCop}$ from \eqref{eq:mlmc_totalerr} for MLMC and as the square root of \eqref{eq:mfmc_mse2} for MFMC. Apart from that, the estimated variance relative to an ordinary Monte Carlo simulation with the same budget is given (rel. MSE). The error estimates for the pressure coefficient $\cp$ are integrated values. For mean and standard deviation, refer to \figref{fig:mfmc_cp_expstd}. 

\begin{table}[htb]

  \centering

  \textbf{(a)}\space\space Lift coefficient $\cl$\\\vspace*{0.1cm}
  \begin{tabular}{ccccc}
    \hline 
               & \bf Mean & \bf StD      & \bf RMSE          & \bf Rel. MSE \\
    \hline 
    \bf Uniced & $0.7783$ & --           & --                & --           \\
    \bf NIPC   & $0.6997$ & $0.0862$     & --                & --           \\
    \bf MLMC   & $0.6795$ & $0.0784$     & $1.14\times10^{-2}$ & $0.520$      \\
    \bf MFMC   & $0.6571$ & $0.0822$     & $9.47\times10^{-3}$ & $0.362$      \\
    \hline 
  \end{tabular}

  \vspace*{0.5cm}
  \textbf{(b)}\space\space Drag coefficient $\cd$\\\vspace*{0.1cm}
  \begin{tabular}{ccccc}
    \hline 
               & \bf Mean & \bf StD      & \bf RMSE          & \bf Rel. MSE \\
    \hline 
    \bf Uniced & $0.0178$ & --           & --                & --           \\ 
    \bf NIPC   & $0.0229$ & $0.0082$     & --                & --           \\ 
    \bf MLMC   & $0.0249$ & $0.0140$     & $1.10\times10^{-3}$ & $0.390$      \\ 
    \bf MFMC   & $0.0278$ & $0.0116$     & $5.01\times10^{-4}$ & $0.081$      \\ 
    \hline 
  \end{tabular}

  \vspace*{0.5cm}
  \textbf{(c)}\space\space Pressure coefficient $\cp$\\\vspace*{0.1cm}
  \begin{tabular}{ccccc}
    \hline 
             & \phantom{\bf Mean} & \phantom{\bf StD}      & \bf RMSE          & \bf Rel. MSE \\
    \hline 
    \bf MLMC &                    &                        & $2.09\times10^{-2}$ & $ 0.503$     \\
    \bf MFMC & \phantom{$0.0000$} &\phantom{$0.0000$}      & $1.46\times10^{-2}$ & $ 0.245$     \\
    \hline 
  \end{tabular}
  \caption{Summary of results and error estimates for icing QoIs}
  \label{tab:icing_results}
\end{table}

The mean of the lift coefficient is reduced substantially compared with the uniced case. The increase in drag is even more severe. Standard deviations are large for the lift, and  for the drag (note that spans between minimum and maximum encountered values are much wider than those given by the standard deviation).

Agreement between the methods in the mean is reasonable. The NIPC method predicts the highest lift and the lowest drag, while the MFMC method predicts the lowest lift and the highest drag. The MLMC method predicts intermediate values in both cases. The lower predicted icing effect of the NIPC method is expected. It is due to the fact that only the first two geometric modes were considered, which leads to a smoother ice shape and a less disturbed flow. The differences between the MLMC and the MFMC methods are owed to sampling errors. The difference is large compared with the estimated sampling standard errors. However, recall that the simulation is optimized for a minimal error of the pressure coefficient $\cp$. The number of samples in both Monte Carlo simulations and the choice of the models in MFMC are thus sub-optimal with respect to the other two QoIs. The presented error \textit{estimates} for $\cl$ and $\cd$, however, express the error that could be achieved with a simulation optimized with resepct to these QoIs. This yields a lower value than the actual stochastic mean of the error. 

Agreement in the standard deviation between the Monte Carlo methods is good. The NIPC method predicts the highest standard deviation in the lift and by far the lowest in the drag. This behavior can again be attributed to the reduced number of considered geometric modes for NIPC: More regular ice shapes are associated with the first two modes. Therefore, there are very few high-drag cases in NIPC, and some even have rather high lift, as opposed to Monte Carlo, where drag is very high in some cases, while lift is almost always reduced. This is further elaborated in the discussion of NIPC response surfaces (\figref{fig:icing_response_surface}).

The relative MSE indicates the fraction of the needed computational budget in comparison with a standard Monte Carlo method with the same accuracy. This is the reciprocal value of both the variance reduction and the speed-up, which are identical here. It is important to note that these values are only estimated. The values range between 0.39 and 0.52 for MLMC and between 0.08 and 0.36 for MFMC. This corresponds to a speed-up of about 2-2.5 for MLMC and 3-12 for MFMC. MFMC outperforms MLMC for all quantities of interest. The difference is largest for the drag coefficient. 

In general, the relative MSE of MLMC and MFMC is determined by the low-fidelity models. If they yield results which are almost identical to the high-fidelity model, but at a much lower computational cost, this leads to a great variance reduction. In the present case, the low and high-fidelity models show considerably different results, which impairs the performance. Further work towards more accurate low-fidelity models such as RANS models could improve speed-up.
The better performance of the MFMC method can be attributed to the additional two-dimensional low-fidelity models and to the possibility of a weighting via the coefficients $\mfmcAl^{\iModel}$.

\subsection{MLMC method-specific quantities}
\label{sec:mlmc_method_results}

\begin{figure}[t!]
  \centering
  \usetikzlibrary{calc}

\begin{tikzpicture}

\def\workwidth{2.65cm}
\def\barwidth{10pt}
\pgfplotstableread{plots/icing/csv_mlopt/mlmc/sigmasq.csv}{\sigmasqtable}
\pgfplotstableread{plots/icing/csv_mlopt/mlmc/mlopt.csv}{\mlopttable}
\pgfplotstableread{plots/icing/csv_mlopt/mlmc/work.csv}{\worktable}

\begin{groupplot}[group style={group name=rzrxcomp, group size=1 by 3,horizontal sep = 1.7cm,  vertical sep = 1.5cm},
width=0.85 \linewidth,
height=0.2 \linewidth,
scale only axis,
]

\nextgroupplot[
scale only axis,
xminorticks=true,
major x tick style = transparent,
major y tick style = transparent,
ybar=2*\pgflinewidth,
bar width=\barwidth,
ymajorgrids = true,
xmode=linear,
ymode=log,
ymin=1,
enlarge x limits=0.20,
yminorticks=false,
xtick pos=right,
xtick={0,1,2},
xticklabels={\tikztitle{$\cl$} (not used),\tikztitle{$\cd$} (not used),\tikztitle{$\cp$ (used)}},
ytick={1,10,100},
yticklabels={$10^{-2}$,$10^{-1}$,$10^{0}$},
ylabel={$\dVar[\Delta\sdqoi^{\levelind}]\,/\,\dVar[\sdqoi^{1}]$},
scatter/position=absolute,
table/x=n,
node near coords style={
    at={(axis cs:\pgfkeysvalueof{/data point/x},1)},
    anchor=east,
    rotate=90,
    color=black,
},
]

\addplot[fill=v6_1,draw=none,nodes near coords=N5-N3]  table [x expr=\coordindex, y index=0] {\sigmasqtable};
\addplot[fill=v6_2,draw=none,nodes near coords=N3-N1]  table [x expr=\coordindex, y index=1] {\sigmasqtable};
\addplot[fill=v6_3,draw=none,nodes near coords=N1   ]  table [x expr=\coordindex, y index=2] {\sigmasqtable};


\nextgroupplot[
scale only axis,
xminorticks=true,
major x tick style = transparent,
major y tick style = transparent,
ybar=2*\pgflinewidth,
bar width=\barwidth,
ymajorgrids = true,
ymode=log,
ymin=1,
ymax=1000,
enlarge x limits=0.20,
yminorticks=false,
ylabel={$\mlopt$},
xtick pos=right,
x tick label style={yshift=-2mm},
xtick=\empty,
ytick={1,10,100,1000,10000},
scatter/position=absolute,
table/x=n,
node near coords style={
    at={(axis cs:\pgfkeysvalueof{/data point/x},1)},
    anchor=east,
    rotate=90,
    color=black,
},
]

\addplot[fill=v6_1,draw=none,nodes near coords=N5-N3]  table [x expr=\coordindex, y index=0] {\mlopttable};
\addplot[fill=v6_2,draw=none,nodes near coords=N3-N1]  table [x expr=\coordindex, y index=1] {\mlopttable};
\addplot[fill=v6_3,draw=none,nodes near coords=N1   ]  table [x expr=\coordindex, y index=2] {\mlopttable};

\end{groupplot}


\begin{axis}[
at={($(rzrxcomp c1r2.south west)+(+0mm,-1.5cm)$)}, 
anchor=north west, 
height=\barwidth,
width=\workwidth,
xbar stacked,
bar width=\barwidth,
xmin = 0, 
xmax = 1,
ymin = -0.1, 
ymax = 0.1,
xtick=\empty,
ytick=\empty,
scale only axis, 
ylabel={work},
y label style={yshift=6mm},
node near coords style={
    at={(axis cs:\pgfkeysvalueof{/data point/y},\pgfkeysvalueof{/pgfplots/xmin})},
    xshift={-3mm},
    anchor=south,
    rotate=90,
    color=black,
},
]
\addplot[fill=v6_1,draw=none,nodes near coords=share] table [y expr=\coordindex, x index=0] {\worktable};
\addplot[fill=v6_2,draw=none] table [y expr=\coordindex, x index=1] {\worktable};
\addplot[fill=v6_3,draw=none] table [y expr=\coordindex, x index=2] {\worktable};
\end{axis}


\begin{axis}[
at={($(rzrxcomp c1r2.south)+(+0mm,-1.5cm)$)}, 
anchor=north, 
height=\barwidth,
width=\workwidth,
xbar stacked,
bar width=\barwidth,
xmin = 0, 
xmax = 1,
ymin = 0.9, 
ymax = 1.1,
xtick=\empty,
ytick=\empty,
scale only axis, 
]
\addplot[fill=v6_1,draw=none] table [y expr=\coordindex, x index=0] {\worktable};
\addplot[fill=v6_2,draw=none] table [y expr=\coordindex, x index=1] {\worktable};
\addplot[fill=v6_3,draw=none] table [y expr=\coordindex, x index=2] {\worktable};
\end{axis}


\begin{axis}[
at={($(rzrxcomp c1r2.south east)+(+0mm,-1.5cm)$)}, 
anchor=north east, 
height=\barwidth,
width=\workwidth,
xbar stacked,
bar width=\barwidth,
xmin = 0, 
xmax = 1,
ymin = 1.9, 
ymax = 2.1,
xtick=\empty,
ytick=\empty,
scale only axis, 
]
\addplot[fill=v6_1,draw=none] table [y expr=\coordindex, x index=0] {\worktable};
\addplot[fill=v6_2,draw=none] table [y expr=\coordindex, x index=1] {\worktable};
\addplot[fill=v6_3,draw=none] table [y expr=\coordindex, x index=2] {\worktable};
\end{axis}


\end{tikzpicture}%
  \caption{Optimal MLMC parameters for different QoIs}
  \label{fig:mlmc_mlopt}
\end{figure}
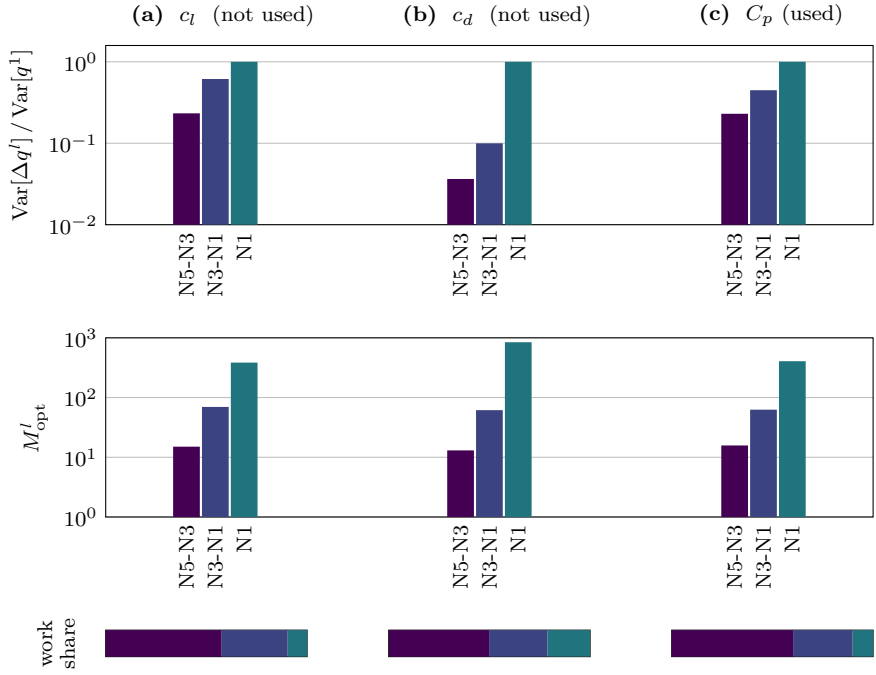

Optimal parameters and related method-specific results of the two Monte Carlo methods are presented in the following, starting with the MLMC method. \figref{fig:mlmc_mlopt} shows the estimated level difference variances $\dVar[\Delta\sdqoi^{\levelind}]$, 
the estimated optimal samples numbers $\mlopt$ and the estimated optimal work shares of the three MLMC levels for the three QoIs. The values are taken after the third iteration, such that the values of $\mlopt$ for $\cp$ reflect the actually computed number of samples. The sorting of the levels is inverted compared with the notation in \secref{sec:mlmc} to ensure comparability to the corresponding \figref{fig:mfmc_mlopt} for MFMC. The level difference variances are normalized with the level variance on the first level.

Level difference variances which decline with increasing resolution (meaning that the baseline model converges) are a condition for MLMC to be useful. This can be confirmed for all three quantities of interest. The decline is strongest for $\cd$ and similar for the two other QoIs, which is in line with the results from \tabref{tab:icing_results}, where the speed-up of MLMC was predicted to be largest for $\cd$ as a QoI, as the drag is easiest to predict even with a coarse grid resolution. 

The estimated optimal sample numbers are rather similar especially for $\cl$ and $\cp$, with more samples on the first level (N1) for $\cd$ and fewer on the third level. This translates to a rather similar optimal share of computational cost for the three QoIs. The finest level needs about half of the computational cost and the coarsest level needs the smallest share. 

\subsection{MFMC method-specific quantities}
\label{sec:mfmc_method_results}

\begin{figure}[t!]
  \centering
  \usetikzlibrary{calc}

\begin{tikzpicture}

\def\workwidth{2.65cm}
\def\barwidth{10pt}
\pgfplotstableread{plots/icing/csv_mlopt/mfmc/rho.csv}{\rhotable}
\pgfplotstableread{plots/icing/csv_mlopt/mfmc/alpha.csv}{\alphatable}
\pgfplotstableread{plots/icing/csv_mlopt/mfmc/mlopt.csv}{\mlopttable}
\pgfplotstableread{plots/icing/csv_mlopt/mfmc/work.csv}{\worktable}

\begin{groupplot}[group style={group name=rzrxcomp, group size=1 by 4,horizontal sep = 1.7cm,  vertical sep = 1.5cm},
width=0.85 \linewidth,
height=0.2 \linewidth,
scale only axis,
]

\nextgroupplot[
scale only axis,
xminorticks=true,
major x tick style = transparent,
major y tick style = transparent,
ybar=2*\pgflinewidth,
bar width=\barwidth,
ymajorgrids = true,
xmode=linear,
ymode=linear,
ymin=0,
ymax=1,
enlarge x limits=0.20,
yminorticks=false,
xtick pos=right,
xtick={0,1,2},
xticklabels={\tikztitle{$\cl$} (not used),\tikztitle{$\cd$} (not used),\tikztitle{$\cp$ (used)}},
ylabel={$\estim{\pears}^{1,\iModel}$},
scatter/position=absolute,
table/x=n,
node near coords style={
    at={(axis cs:\pgfkeysvalueof{/data point/x},\pgfkeysvalueof{/pgfplots/ymin})},
    anchor=east,
    rotate=90,
    color=black,
},
]

  \addplot[fill=v6_1,draw=none,nodes near coords=HF   ]  table [x expr=\coordindex, y expr=sqrt(\thisrowno{0})] {\rhotable};
  \addplot[fill=v6_2,draw=none,nodes near coords=3D N3]  table [x expr=\coordindex, y expr=sqrt(\thisrowno{1})] {\rhotable};
  \addplot[fill=v6_3,draw=none,nodes near coords=3D N1]  table [x expr=\coordindex, y expr=sqrt(\thisrowno{2})] {\rhotable};
  \addplot[fill=v6_4,draw=none,nodes near coords=2D N5]  table [x expr=\coordindex, y expr=sqrt(\thisrowno{3})] {\rhotable};
  \addplot[fill=v6_5,draw=none,nodes near coords=2D N3]  table [x expr=\coordindex, y expr=sqrt(\thisrowno{4})] {\rhotable};
  \addplot[fill=v6_6,draw=none,nodes near coords=2D N1]  table [x expr=\coordindex, y expr=sqrt(\thisrowno{5})] {\rhotable};


\nextgroupplot[
scale only axis,
xminorticks=true,
major x tick style = transparent,
major y tick style = transparent,
ybar=2*\pgflinewidth,
bar width=\barwidth,
ymajorgrids = true,
ymode=log,
ymin=1,
ymax=10000,
enlarge x limits=0.20,
yminorticks=false,
ylabel={$\mlopt$},
xtick pos=right,
x tick label style={yshift=-2mm},
xtick=\empty,
ytick={1,10,100,1000,10000},
scatter/position=absolute,
table/x=n,
node near coords style={
    at={(axis cs:\pgfkeysvalueof{/data point/x},1)},
    anchor=east,
    rotate=90,
    color=black,
},
]

\addplot[fill=v6_1,draw=none,nodes near coords=HF   ]  table [x expr=\coordindex, y index=0] {\mlopttable};
\addplot[fill=v6_2,draw=none,nodes near coords=3D N3]  table [x expr=\coordindex, y index=1] {\mlopttable};
\addplot[fill=v6_3,draw=none,nodes near coords=3D N1]  table [x expr=\coordindex, y index=2] {\mlopttable};
\addplot[fill=v6_4,draw=none,nodes near coords=2D N5]  table [x expr=\coordindex, y index=3] {\mlopttable};
\addplot[fill=v6_5,draw=none,nodes near coords=2D N3]  table [x expr=\coordindex, y index=4] {\mlopttable};
\addplot[fill=v6_6,draw=none,nodes near coords=2D N1]  table [x expr=\coordindex, y index=5] {\mlopttable};

\nextgroupplot[
scale only axis,
xminorticks=true,
major x tick style = transparent,
major y tick style = transparent,
ybar=2*\pgflinewidth,
bar width=\barwidth,
ymajorgrids = true,
xmode=linear,
ymode=linear,
ymin=0,
ymax=1,
enlarge x limits=0.20,
yminorticks=true,
ylabel={$\mfmcAl^{\iModel}$},
xtick pos=right,
x tick label style={yshift=-2mm},
xtick=\empty,
scatter/position=absolute,
table/x=n,
node near coords style={
    at={(axis cs:\pgfkeysvalueof{/data point/x},\pgfkeysvalueof{/pgfplots/ymin})},
    anchor=east,
    rotate=90,
    color=black,
},
]

\addplot[fill=v6_1,draw=none,nodes near coords=HF   ]  table [x expr=\coordindex, y index=0] {\alphatable};
\addplot[fill=v6_2,draw=none,nodes near coords=3D N3]  table [x expr=\coordindex, y index=1] {\alphatable};
\addplot[fill=v6_3,draw=none,nodes near coords=3D N1]  table [x expr=\coordindex, y index=2] {\alphatable};
\addplot[fill=v6_4,draw=none,nodes near coords=2D N5]  table [x expr=\coordindex, y index=3] {\alphatable};
\addplot[fill=v6_5,draw=none,nodes near coords=2D N3]  table [x expr=\coordindex, y index=4] {\alphatable};
\addplot[fill=v6_6,draw=none,nodes near coords=2D N1]  table [x expr=\coordindex, y index=5] {\alphatable};

\end{groupplot}


\begin{axis}[
at={($(rzrxcomp c1r3.south west)+(+0mm,-1.5cm)$)}, 
anchor=north west, 
height=\barwidth,
width=\workwidth,
xbar stacked,
bar width=\barwidth,
xmin = 0, 
xmax = 1,
ymin = -0.1, 
ymax = 0.1,
xtick=\empty,
ytick=\empty,
scale only axis, 
ylabel={work},
y label style={yshift=6mm},
node near coords style={
    at={(axis cs:\pgfkeysvalueof{/data point/y},\pgfkeysvalueof{/pgfplots/xmin})},
    xshift={-3mm},
    anchor=south,
    rotate=90,
    color=black,
},
]
\addplot[fill=v6_1,draw=none,nodes near coords=share] table [y expr=\coordindex, x index=0] {\worktable};
\addplot[fill=v6_2,draw=none] table [y expr=\coordindex, x index=1] {\worktable};
\addplot[fill=v6_3,draw=none] table [y expr=\coordindex, x index=2] {\worktable};
\addplot[fill=v6_4,draw=none] table [y expr=\coordindex, x index=3] {\worktable};
\addplot[fill=v6_5,draw=none] table [y expr=\coordindex, x index=4] {\worktable};
\addplot[fill=v6_6,draw=none] table [y expr=\coordindex, x index=5] {\worktable};
\end{axis}


\begin{axis}[
at={($(rzrxcomp c1r3.south)+(+0mm,-1.5cm)$)}, 
anchor=north, 
height=\barwidth,
width=\workwidth,
xbar stacked,
bar width=\barwidth,
xmin = 0, 
xmax = 1,
ymin = 0.9, 
ymax = 1.1,
xtick=\empty,
ytick=\empty,
scale only axis, 
]
\addplot[fill=v6_1,draw=none] table [y expr=\coordindex, x index=0] {\worktable};
\addplot[fill=v6_2,draw=none] table [y expr=\coordindex, x index=1] {\worktable};
\addplot[fill=v6_3,draw=none] table [y expr=\coordindex, x index=2] {\worktable};
\addplot[fill=v6_4,draw=none] table [y expr=\coordindex, x index=3] {\worktable};
\addplot[fill=v6_5,draw=none] table [y expr=\coordindex, x index=4] {\worktable};
\addplot[fill=v6_6,draw=none] table [y expr=\coordindex, x index=5] {\worktable};
\end{axis}


\begin{axis}[
at={($(rzrxcomp c1r3.south east)+(+0mm,-1.5cm)$)}, 
anchor=north east, 
height=\barwidth,
width=\workwidth,
xbar stacked,
bar width=\barwidth,
xmin = 0, 
xmax = 1,
ymin = 1.9, 
ymax = 2.1,
xtick=\empty,
ytick=\empty,
scale only axis, 
]
\addplot[fill=v6_1,draw=none] table [y expr=\coordindex, x index=0] {\worktable};
\addplot[fill=v6_2,draw=none] table [y expr=\coordindex, x index=1] {\worktable};
\addplot[fill=v6_3,draw=none] table [y expr=\coordindex, x index=2] {\worktable};
\addplot[fill=v6_4,draw=none] table [y expr=\coordindex, x index=3] {\worktable};
\addplot[fill=v6_5,draw=none] table [y expr=\coordindex, x index=4] {\worktable};
\addplot[fill=v6_6,draw=none] table [y expr=\coordindex, x index=5] {\worktable};
\end{axis}


\end{tikzpicture}%
  \caption{Optimal MFMC parameters for different QoIs; Labels of the form N1 3D refer to the DG polynomial degree $\ndg=1$ and the physical dimensionality.}
  \label{fig:mfmc_mlopt}
\end{figure}
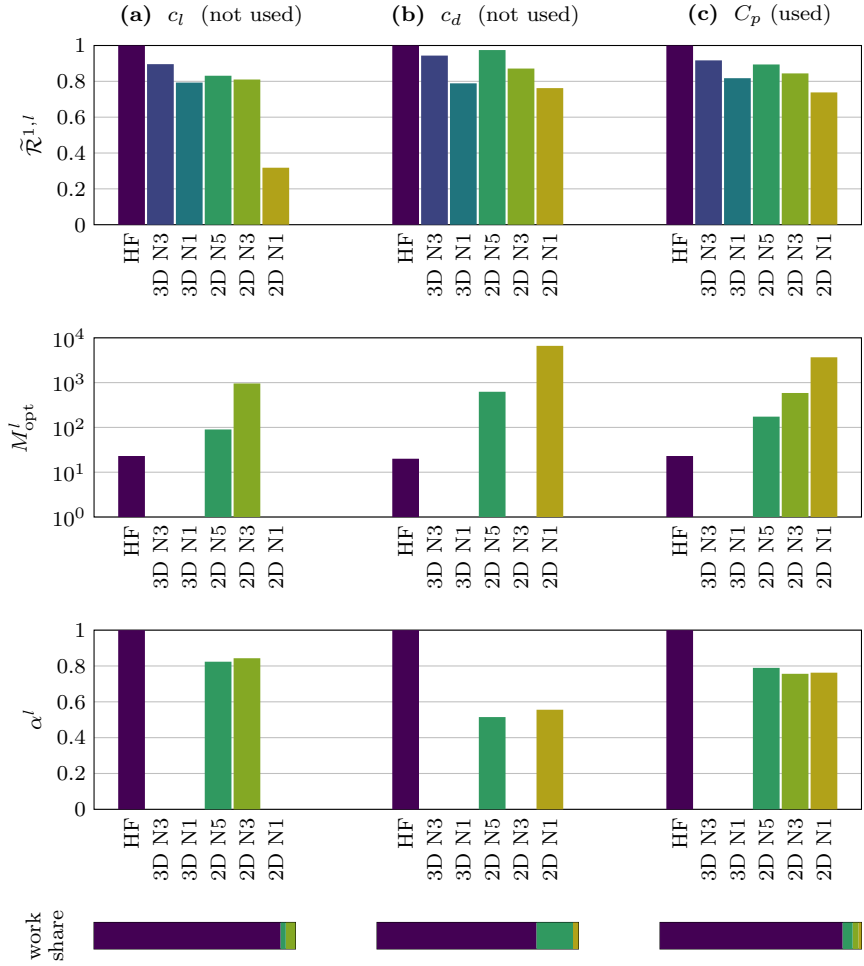

\figref{fig:mfmc_mlopt} shows estimates for optimal MFMC parameters for the three QoIs.
The first row shows the Pearson correlation coefficient $\estim{\pears}^{1,\iModel}$ between high and low-fidelity models. 
It has values above 0.75 except for the two-dimensional model with $\ndg=1$, which has a low correlation with the high-fidelity model. Surprisingly, the two-dimensional model with $\ndg=5$ shows similar or even superior correlation values compared with the three-dimensional low-fidelity models despite the absence of turbulence in two-dimensional simulations.
Correlations are lower than the ones found in the literature (cf. \cite{Peherstorfer2016,PyMLMCSukys2017}). Possible reasons include the complexity and high sensitivity of the investigated problem, the relatively low grid resolutions, the residual randomness from turbulent fluctuations, and the design and choice of low-fidelity models.

The second row shows the optimal sample numbers estimated for each QoI. In MFMC, an optimal subset of models is selected, while the other low-fidelity models are discarded. The two three-dimensional low-fidelity models are discarded for all QoIs. An optimal configuration for the prediction of the lift coefficient $\cl$ further excludes the two-dimensional model with $\ndg=1$, while for the drag coefficient $\cd$, excluding the two-dimensional model with $\ndg=3$ yields optimal results. Recall that the selection of low-fidelity models and number of samples is optimized with respect to the pressure coefficient $\cp$. This means that only the choice of models and sample numbers for $\cp$ are actually used. So all two-dimensional low-fidelity models were included in the MFMC simulation. Several thousand samples with the lowest fidelity model are optimal for $\cd$ and $\cp$, which is more than in the MLMC method. The two-dimensional models, which were not included in MLMC, are much cheaper to evaluate than the three-dimensional models, so that a higher sample number can be computed.

The coefficients $\mfmcAl^{\iModel}$ are shown in the third row. They are rather similar at values around 0.8 for the QoIs $\cl$ and $\cp$, but significantly lower for $\cd$. The coefficients are scaled by the ratio of variances between the high and low fidelity model. Drag is overestimated in the low-fidelity models due to numerical viscosity, which leads to a higher variance and thus lower $\mfmcAl^{\iModel}$. This possibility to account for systematic differences in QoI variance is an advantage of MFMC over MLMC.

As shown in the last row of the figure, the bulk of the computational budget is used for high-fidelity samples. This is also due to the much lower per-sample cost of the two-dimensional low-fidelity models.

\subsection{Sample flow fields and response surfaces}
\label{sec:flowfields_results}

\begin{figure}[t!]
  \centering
  \setlength{\fboxsep}{0pt}%
  \setlength{\fboxrule}{0.5pt}%

  \vspace*{-1mm}
  \tikztitle{Uniced}
  \fbox{\includegraphics[width=10.2cm, trim=35mm 135mm 10mm 135mm,clip=true]{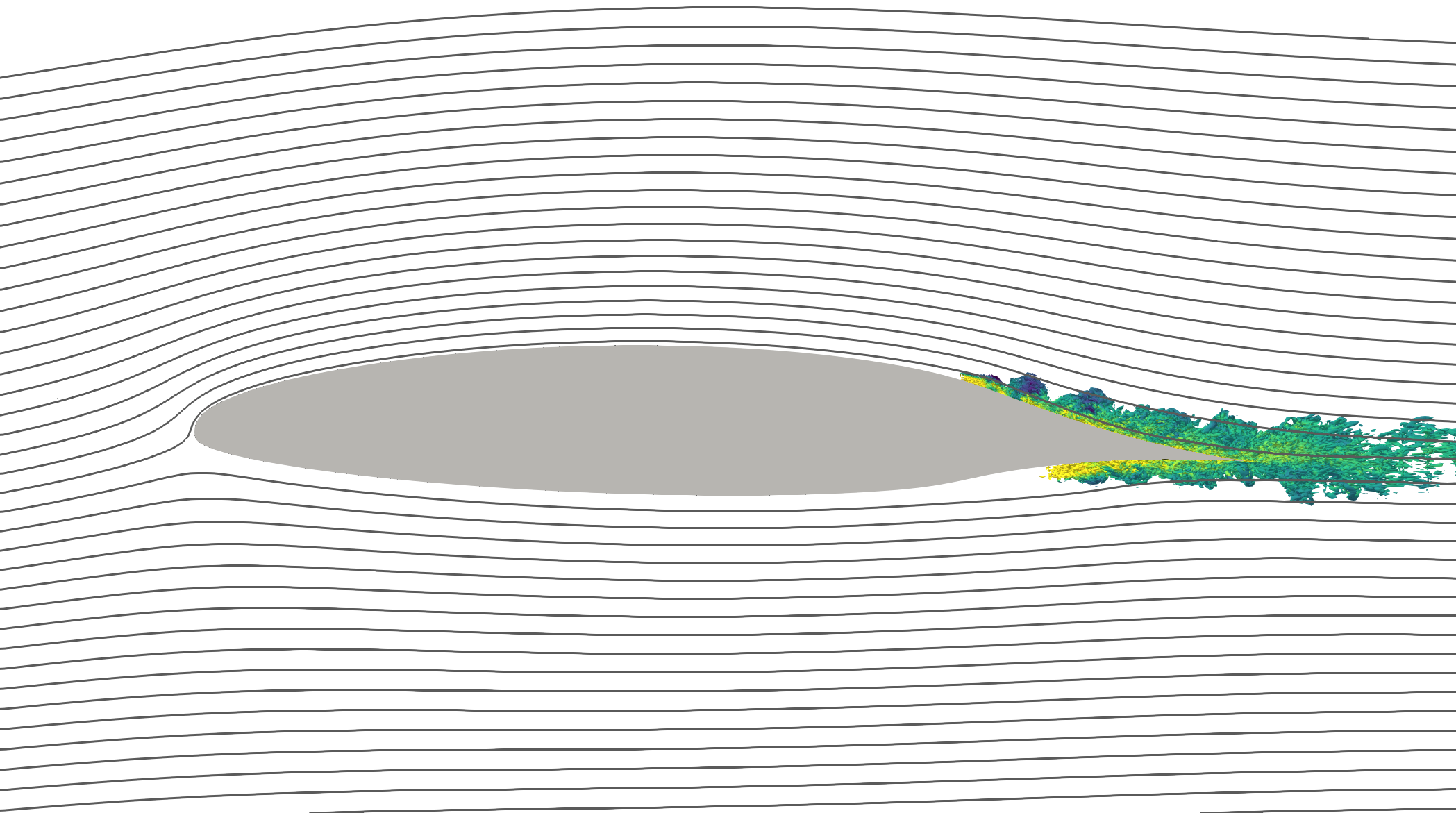}}

  \vspace*{3mm}
  \tikztitle{Iced Monte Carlo samples}
  \fbox{\includegraphics[width=10.2cm, trim=35mm 135mm 10mm 135mm,clip=true]{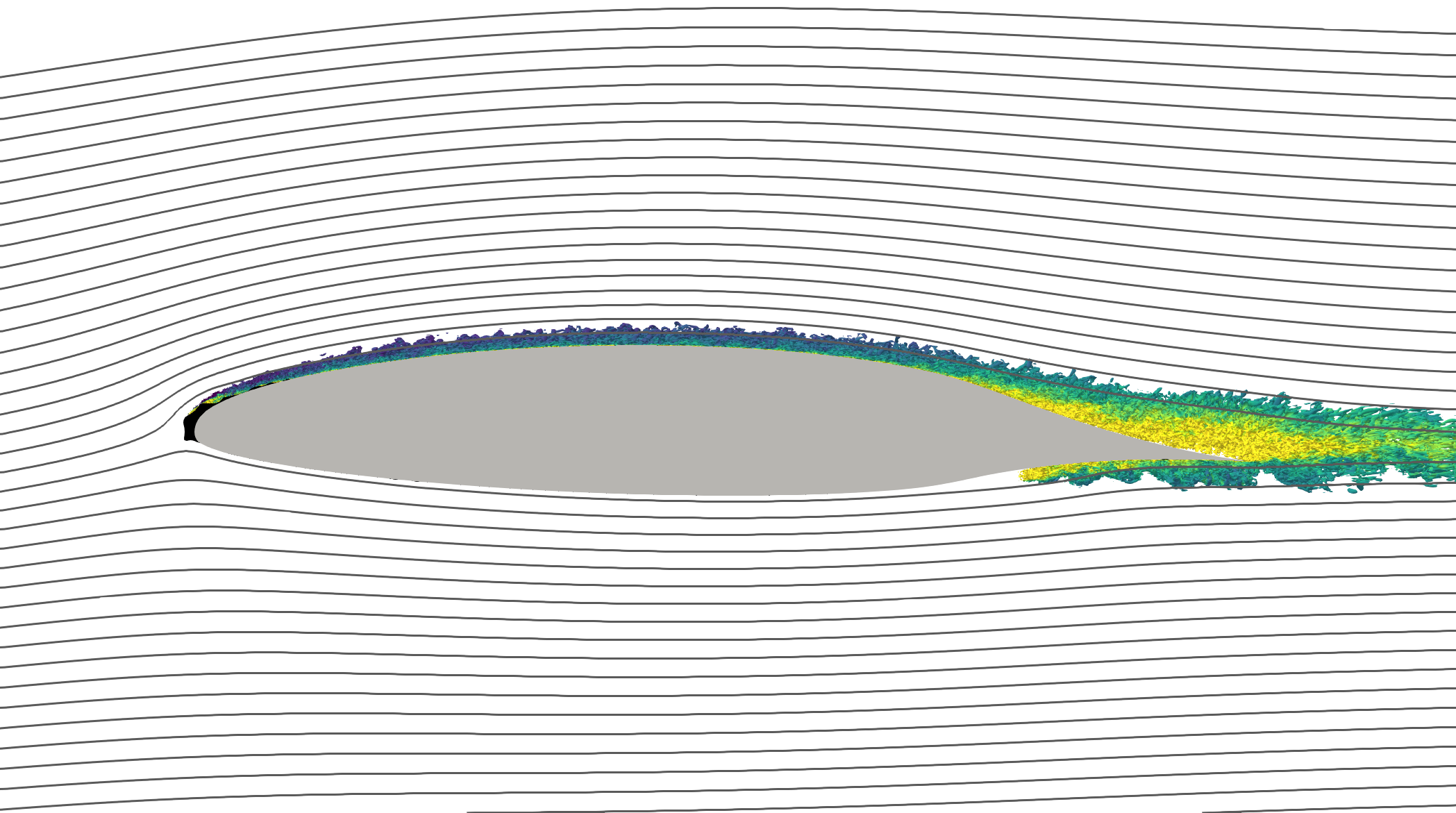}}
  
  \vspace*{2.5mm}
  \fbox{\includegraphics[width=10.2cm, trim=35mm 135mm 10mm 135mm,clip=true]{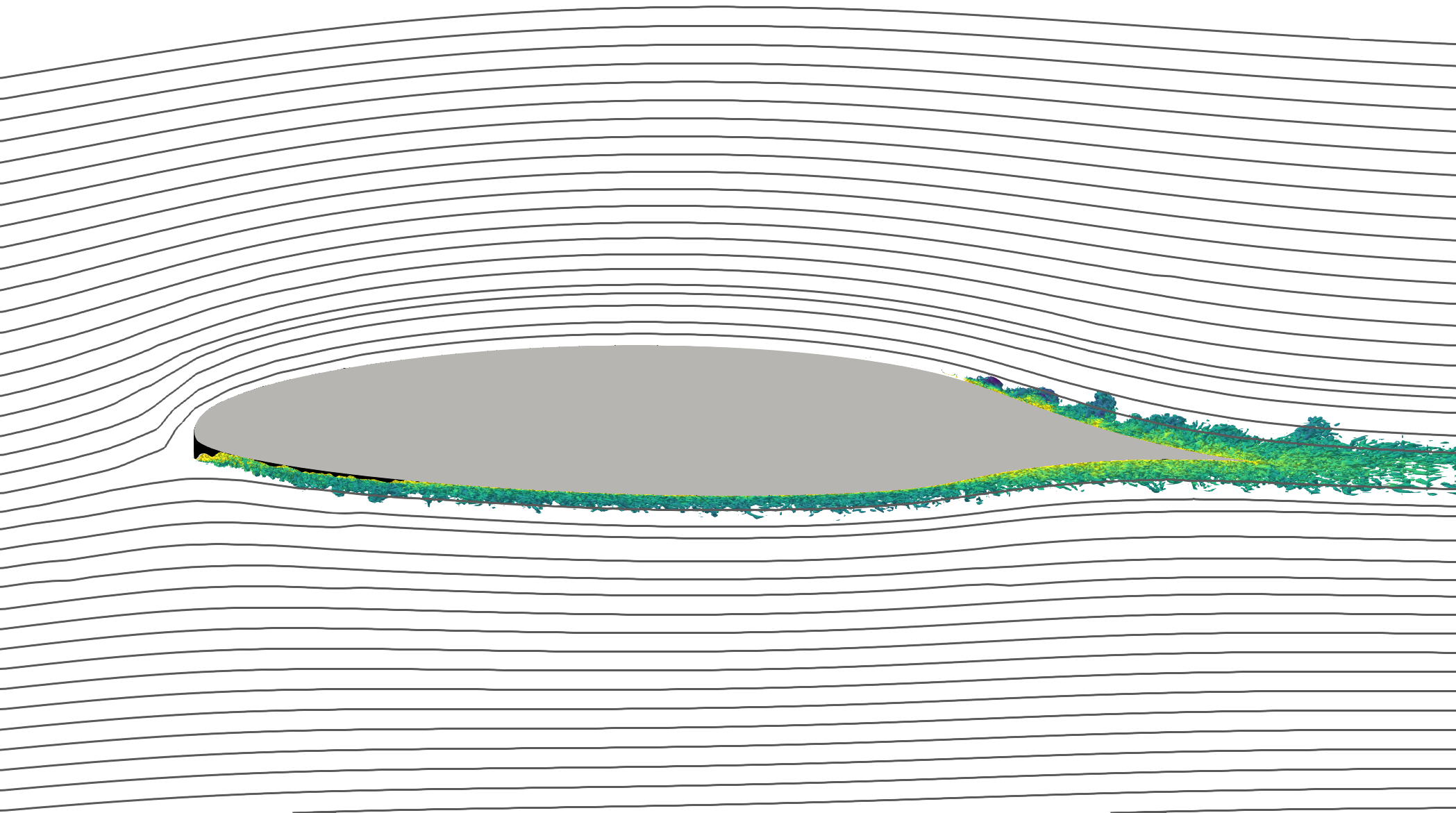}}
  
  \vspace*{2.5mm}
  \fbox{\includegraphics[width=10.2cm, trim=35mm 135mm 10mm 135mm,clip=true]{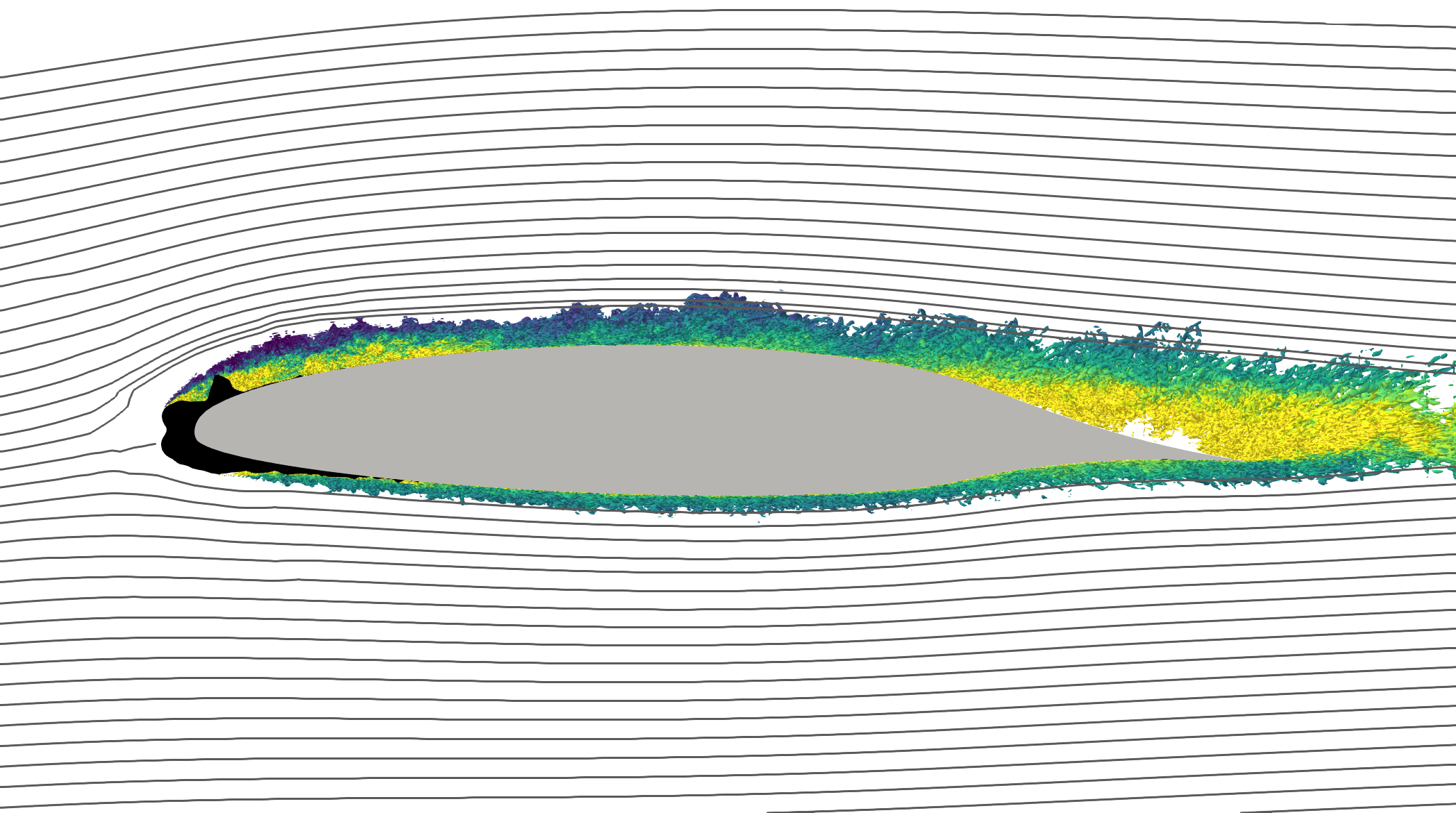}}
  
  \vspace*{2.5mm}
  \fbox{\includegraphics[width=10.2cm, trim=35mm 135mm 10mm 135mm,clip=true]{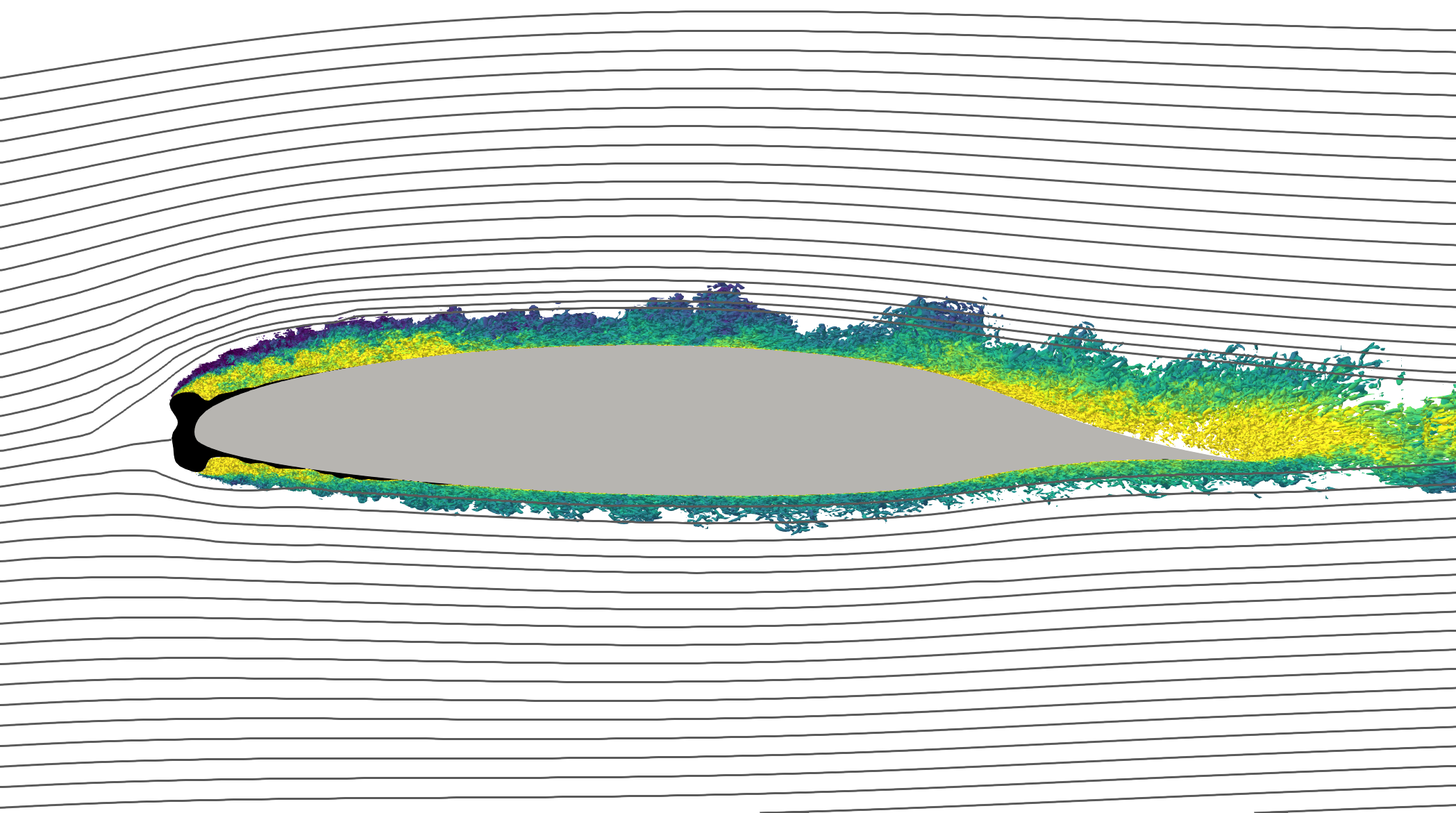}}

  \vspace*{1.5mm}
  \begin{tikzpicture}
    \begin{axis}[
        ytick=\empty,
        scale only axis,
        enlargelimits=false,
        xtick align=outside,
        xtick pos=left,
        width = 3cm,
        height = 0.3cm,
        xlabel={$\left|v\right|$},
        ]
      \addplot graphics[xmin=0.2,xmax=1.6,ymin=0.,ymax=1.0] {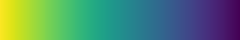};
      \draw (0.2,0) rectangle (1.6,1.0);
    \end{axis}
  \end{tikzpicture}
  \vspace*{-1mm}
  \caption[Uniced airfoil flow field compared with four iced Monte Carlo samples]{Uniced airfoil flow field (top) compared with four iced Monte Carlo samples (bottom); Ice shapes are colored black, instantaneous Q-criterion isosurfaces are colored by velocity magnitude, streamlines of time averaged airfoil are gray.}
  \label{fig:icing_mc_fieldsol}
\end{figure}

Ice shapes and instantaneous flow fields of the uniced airfoil and four iced Monte Carlo samples are shown in \figref{fig:icing_mc_fieldsol}. The ice shapes are colored black. The Monte Carlo samples are the same as those shown in Figures \ref{fig:icing_grid} (a) and \ref{fig:pca} (e), where the ice shapes can be seen in more detail. Isosurfaces of the Q-criterion colored by velocity magnitude visualize instantaneous boundary layer snapshots. Time-averaged streamlines indicate separation regions. 

The flow around the uniced airfoil stays laminar until about 75 \% chord length, where transition occurs on the suction side. Transition on the pressure side is slightly further downstream. Laminar separation bubbles with turbulent re-attachment are present on both sides. The one on the suction side is very small. The one on the pressure side is much larger.

The Monte Carlo sample flow fields demonstrate the large influence of the ice geometry. In the first sample, the ice triggers transition at the leading edge on the suction side, in the second sample on the pressure side, and in the last two samples on both sides. The state of the turbulent boundary layers shows large variations. A rather regular and thin boundary layer can be seen in the first two samples, while the second two show a highly disturbed flow with large vortices, indicated by large fluctuations in the thickness of the boundary layer visualization. Note that the thickness of the region where the Q-criterion exceeds certain values is only an approximate indicator for the turbulent boundary layer thickness. Regions of separated flow are present behind the ice horns. They are largest on the suction side of the third sample and on both sides of the fourth sample. The disturbed flow on the suction side leads to separation bubbles in the aft half of the airfoil. They are largest in the last two samples, but still considerable in the first one. The laminar separation bubble on the pressure side is diminished in the first sample and absent due to the leading-edge transition in the last three.

\begin{figure}[t!]
  \centering
  \input{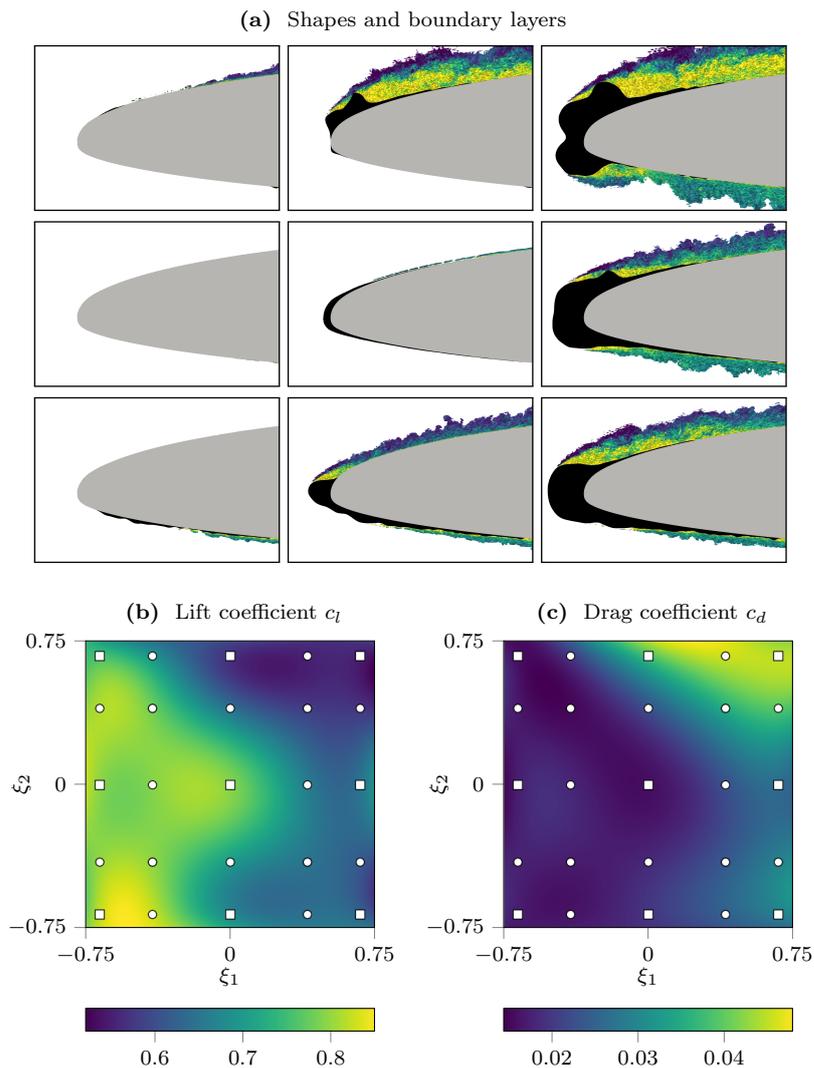}
  \caption[NIPC field solutions and response surfaces]{NIPC field solutions and response surfaces. Flow fields are visualized as in \figref{fig:icing_mc_fieldsol}. The white dots in the response surfaces indicate NIPC quadrature points. The squares indicate the simulations shown on the top, with the same alignment along $\xicmp_1$ and $\xicmp_2$.}
  \label{fig:icing_response_surface}
\end{figure}

The same visualization of the ice shapes and instantaneous flow fields is shown for the leading edge region of nine NIPC quadrature point simulations in \figref{fig:icing_response_surface} (a). In each stochastic dimension (each of which corresponds to one uncertain parameter, i.e. one geometric mode), five quadrature points were used. The figure shows the first, the third and the fifth in each dimension. The first parameter $\xicmp_1$ is increased from left to right, the second one $\xicmp_2$ is increased from bottom to top. 

The NIPC method approximates the response of the QoI to the uncertain input as a polynomial. The polynomial response surface is shown for $\cl$ in \figref{fig:icing_response_surface} (b) and for $\cd$ in \figref{fig:icing_response_surface} (c). The color represents the local value of $\cl$ and $\cd$. The white dots in the response surface indicate the positions of the quadrature points (i.e. the simulations). The squares indicate the subset of samples shown in \figref{fig:icing_response_surface} (a).

The amount of icing is mainly governed by the first mode, where the ice volume increases with increasing $\xicmp_1$. The three samples on the left show almost no ice accretion. The second mode determines whether the icing is located more towards the pressure side (low $\xicmp_2$) or on the suction side (high $\xicmp_2$). Due to the missing high-frequency principal component modes, the ice shapes are rounder than the ones of the four Monte Carlo samples shown in the previous figure. Only two samples (the top center and the top right) show a large vertical extension of the ice horns, which is expected to have the largest aerodynamic effect.

Leading-edge transition to turbulent flow occurs in most of the samples. Separated flow behind the ice (indicated by yellow regions) can be seen in many. The size of the separation regions and turbulent boundary layer thickness, especially on the suction side, are expected to be an indicator of the aerodynamic impact of the ice. 

The response surfaces for $\cl$ and $\cd$ confirm this hypothesis. The lift coefficient is largest for the samples on the left (with almost no separation) and lowest for the top right samples with large separation regions on the suction side. In the drag response surface, a large aerodynamic impact causes an increase in $\cd$.

The highest lift among the NIPC sample simulations even slightly exceeds that on the uniced airfoil ($\cl\approx0.82$ versus $\cl\approx0.78$ on the uniced airfoil). The lowest drag among the NIPC samples is slightly lower than that on the uniced airfoil ($\cd\approx0.016$ versus $\cd\approx0.018$). This suggests that the tripped boundary layer has a positive effect on aerodynamic performance due to the reduced or removed separation bubbles. The airfoil is primarily designed for higher Reynolds numbers of approximately $10^7$. The design objectives include the use with lower Reynolds numbers, but the use of boundary layer trips is specifically recommended for this case in the airfoil specification \cite{Viken1987}. A small positive effect of light icing is therefore plausible. 



The ratio of maximal to minimal lift in the response surface is approximately 1.6, while the ratio of the maximal to minimal drag is approximately 3.3. 
The large ranges of lift and drag emphasize once more the strong uncertain effect of the investigated icing on aerodynamic performance. 

\section{Conclusion}
\label{sec:conclusion}

The aerodynamic effects of uncertain iced airfoil shapes were investigated using a high-order computational fluid dynamics model and different uncertainty quantification methods in a data-integrated approach.

As a baseline CFD model, a discretization of the compressible Navier-Stokes equations based on the discontinuous Galerkin spectral element method was used. Three UQ methods were compared: The non-intrusive polynomial chaos (NIPC) method, the multilevel Monte Carlo (MLMC) method, and the multifidelity Monte Carlo (MFMC) method.

In a data-driven approach, the uncertain input parameters were generated from a set of ice shapes measured in wind tunnel experiments. To this end, a parametrization based on a signed-distance function was introduced and a principal component analysis was performed on the parametrized data. As high sample numbers in Monte Carlo methods require the complete automation of the baseline simulation tool chain, an automated structured boundary layer grid generator was developed to account for the varying computational domains. A wall-resolved LES setup was presented for the non-intrusive sample simulations. Simulations with the three considered non-intrusive methods were compared. Lift coefficient, drag coefficient and pressure coefficient were chosen as quantities of interest (QoIs). 
Lower polynomial degrees of the DG method on the same grid were used as lower fidelity models for MLMC and MFMC. In MFMC, laminar two-dimensional models were additionally used. For the NIPC simulation, only two uncertain geometric modes were used as input, as the cost of NIPC simulations increases exponentially with the number of uncertain parameters. 

Results showed a substantial influence of the icing on the lift, which was on average decreased, and an even larger effect on the drag, which was on average increased. The icing triggered in many cases leading-edge transition on the pressure side, suction side, or both. Some ice shapes included leading-edge horns, behind which separation bubbles formed. The changes in the boundary layer due to the icing also altered the flow in the aft region of the airfoil, sometimes causing large turbulent separation bubbles. In some cases, lift-increasing and drag-reducing effects of the ice were observed, where the ice acted as a boundary layer trip and prevented laminar separation. 
Agreement of the methods in the pressure coefficient, which was the optimized QoI for MLMC and MFMC, was very good. Agreement in the lift and drag coefficients, which the Monte Carlo methods' parameters were not optimized for, was satisfactory. The NIPC method predicted a higher mean lift and lower mean drag, which was owed to the reduced number of uncertain modes, which resulted in more rounded ice shapes and a less disturbed flow. 
The speed-up (equivalent to variance reduction) compared with standard Monte Carlo ranged from 2-2.5 for MLMC and from 3-12 for MFMC, depending on the QoI. The superior performance of MFMC in comparison with MLMC was likely due to the additional two-dimensional models and the additional weighting coefficients in MFMC. The speed-up is determined by the accuracy and per-sample cost of the low-fidelity models. As the QoIs of the considered problem are very sensitive to the model setup, all of the presented low-fidelity models in this study showed substantially different results compared with the high-fidelity model, such that a near-perfect correlation with the high-fidelity model as presented in other works in the literature was not achieved. Response surfaces of the NIPC method were shown for the lift and drag coefficients. They revealed the non-linear dependency of lift and drag on the investigated two geometric modes. They also allowed further insight into the aerodynamic impact of specific ice shapes, such as the positive effect of certain forms of light icing on lift or drag.\\

Future work can focus on different aspects: 
The setup of the baseline CFD model can be further refined. This includes a finer near-wall resolution (requiring a greater computational budget) and validation with experiments.
Grid convergence entails that a finer resolution improves correlation between high-fidelity and low-fidelity models, so that the cost of the UQ simulation only increases under-proportionally with the cost of a high-fidelity sample. For both considered applications, a longer averaging time period for the quantities of interest can further reduce the impact of pseudo-random turbulent fluctuations.

To improve the stochastic accuracy, more accurate and/or cheaper low-fidelity models can be employed for MFMC. This includes, for example, RANS simulations or simulations with a panel method. Surrogate models, where a response surface is constructed based on evaluations of another model, have a large potential in this regard. The existing NIPC interpolation can be used, for example, as such a surrogate model with no significant additional computational cost, if it is assumed to be constant in the omitted stochastic dimensions. Different surrogate models, for instance based on Kriging, might yield even better results. More generally, the employed  MLMC and MFMC are just two variance reduction techniques. Other approaches such as the multilevel-multifidelity Monte Carlo method can be tested. Latin hypercube sampling is an entirely different, but nonetheless promising approach, possibly in combination with the other variance reduction techniques. In NIPC, more uncertain parameters can be considered when using a nested sparse grid and/or a lower, potentially anisotropic stochastic polynomial degree, which can be adaptively increased. This may, but does not necessarily improve the computational efficiency.

For the prediction of the impact of airfoil icing, the set of uncertain input parameters has potential for further improvement by basing it on a larger experimental data set and using non-linear dimensionality reduction techniques instead of PCA. Alternatively, the ice shapes can be gained from computational modeling of the accretion process under uncertain external conditions such as temperature and droplet size.

Beyond airfoil icing, UQ studies of different applications can be carried out efficiently and with comparatively little additional effort, as the necessary software framework is now in place and the strengths and weaknesses of the methods are known.

\begin{acknowledgements}
The authors would like to thank for funding by the Friedrich und Elisabeth Boysen Stiftung via the project ``BOY-143'' and by the Deutsche Forschungsgemeinschaft (DFG, German Research Foundation) under Germany’s Excellence Strategy -EXC 2075 – 390740016.
The authors would also like to thank the High-Performance Computing Center Stuttgart (HLRS) for the provided computing resources through the project ``SEAL''.
\end{acknowledgements}

\section*{Conflict of interest}

The authors declare that they have no conflict of interest.

\appendix
\section{Structured two-dimensional boundary layer grid generation algorithm}
\label{sec:grid_appendix}

An algorithm is qualitatively described which is used for the generation of structured, two-dimensional C-grids for iced airfoils in \secref{sec:grid_generation}. As a first step, a structured grid of the uniced airfoil is generated. The wall-parallel spacing is finer near the leading edge and near the trailing edge than at the rest of the airfoil. A normal extrusion is used in wall-normal direction with an exponential stretching, which ensures a finer spacing near the wall. The grid points of the outer grid layers are re-distributed along the wall-parallel coordinate to achieve a finer wall-parallel spacing around the leading edge also in the outer layers, where the sharp radius would otherwise lead to very large elements. 

Next, the surface of the iced airfoil is prepared. The random ice geometry and the clean airfoil are intersected and the outermost contour of both is used. The wall-parallel spacing on the surface is finer in regions where the ice shape is convex and sparser where it is concave. A slight surface smoothing is applied. The effects are shown and discussed in \secref{sec:grid_generation}. 

An non-smoothed volume grid for the iced airfoil is defined. This grid contains negative elements and is only auxiliary. For its generation, the displacement vector between the iced and the uniced surface grid points is calculated. This offset is then multiplied with a linear wall-normal blending factor which is one at the surface and zero at the outermost grid layer. The result is added to the uniced grid to yield the auxiliary iced grid. 

In order to achieve a valid grid, smoothing is applied to the grid. The smoothing algorithm is discussed below. It oscillates and crashes on low-quality grids, which are far from the equilibrium of the smoothing algorithm. An iterative algorithm is therefore used: Starting with the uniced grid, the displacement towards the iced auxiliary grid is divided into small steps. After each step, smoothing is applied. This ensures a high-quality grid close to the smoothing equilibrium throughout the whole transition from uniced to iced outline. After this displacement phase, the smoothing runs until it converges towards a steady state. 

The smoothing is now discussed. It is based on Laplace smoothing, where the position of each grid point is replaced with the average position of its four direct neighbors.
In wall-normal direction, a weighting of the inner and outer neighbors preserves the grid stretching. 
Laplace smoothing yields high-quality grids in many regions, but fails in others. For example, along sharp convex curves of the ice shapes, negative cells can occur. 

The Laplace smoothing is therefore extended by several terms, which ensure different aspects of grid quality: 
\begin{itemize}
  \item Grid points are moved away from very small elements to ensure a minimum element size. In wall-normal direction, this movement is always directed outwards. 
  \item Grid points are moved to reduce distortions, i.e. to avoid large deviations from $90^{\circ}$ angles.
  \item Grid points are moved if the size ratio of two neighboring elements is very large.
\end{itemize}
For each of the criteria, a factor for the magnitude of displacement is calculated, which depends on how severely the respective grid quality constraint is violated. The magnitude of each term is zero if the respective grid quality metric is good. 

The Laplace smoothing and the three criteria above each create a displacement vector for each grid point. 
These displacement vectors are summed. In unproblematic regions, only the Laplace smoothing term is non-zero. If the resulting vector is too large, it is trimmed to avoid oscillations. 

The first grid layer next to the wall is not treated with the smoothing algorithm. Instead, wall-normal extrusion with a prescribed layer thickness is enforced here. 

Results of the algorithm are shown in \figref{fig:icing_grid} (a) in \secref{sec:grid_generation}.

\bibliographystyle{spmpsci}      
\bibliography{bibliography.bib}   

\end{document}